\documentclass[usenatbib]{mn2e}
\usepackage{graphicx}
\usepackage{amssymb}
\usepackage{amsmath}
\usepackage{times}
\usepackage{aas_macros}

\bibpunct[,]{(}{)}{;}{a}{}{,}

\begin{document}

\title[Multi-fractal and Lacunarity in Millennium DM haloes]{Millennium Simulation Dark Matter Haloes: Multi-fractal and Lacunarity analysis with homogeneity transition}
\author[C. A. Chac\'on-Cardona and R. A.  Casas-Miranda]{C\'esar. A. Chac\'on-Cardona$^{1,2}$ and Rigoberto A. Casas-Miranda$^1$\\
		$^1$Departamento de F\'{\i}sica, Universidad Nacional de Colombia\\
		$^2$Facultad Tecnol\'ogica, Universidad Distrital Francisco Jos\'e de Caldas\\}
		
\date{Accepted ---. Received ---; in original form ---}

\pagerange{\pageref{firstpage}--\pageref{lastpage}} \pubyear{2011}

\maketitle
\label{firstpage}

\begin{abstract}

We investigate from the fractal viewpoint the way in which the dark matter is grouped at $z = 0$ in the Millennium dark matter cosmological simulation. The determination of the cross to homogeneity in the Millennium Simulation data is described from the behaviour of the fractal dimension and the lacunarity. 
We use the sliding window technique to calculate the fractal mass-radius dimension, the pre-factor $F$ and the lacunarity of this fractal relation.  Besides, we determinate the multi-fractal dimension and the lacunarity spectrum, including their dependence with radial distance. This calculations show a radial distance dependency of all the fractal quantities, with heterogeneity clustering of dark matter haloes up to depths of 100 Mpc/h. 
The dark matter haloes clustering in the Millennium Simulation shows a radial distance dependency, with two regions clearly defined. The lacunarity spectrum for values of the structure parameter $q \geqslant 1$ shows regions with relative maxima, revealing the formation of clusters and voids in the dark matter haloes distribution. With the use of the multi-fractal dimension and the lacunarity spectrum, the transition to homogeneity at depths between 100 Mpc/h and 120 Mpc/h for the Millennium Simulation dark matter haloes is detected.

\end{abstract}


\section{Introduction}

The current paradigm,  which is the basis of the standard cosmological model, states that the large-scale universe is homogeneous and isotropic, i.e., there is no privileged place or direction in the universe  \citep{2002SPIE.4847...86M} in a way that the observed inhomogeneities are local in character and they should vanish at sufficiently large scales  \citep{peacock1999cosmological, longair2008galaxy}. This principle is based mainly on the observations of the cosmic microwave radiation whose isotropy is one part in a hundred thousand \citep{2005astro.ph..5185B},  and philosophical considerations: the observed universe should be the same for any observer, the Copernican Principle.

Despite the success of the physical models constructed from this cosmological principle, some researchers now are debating heavily whether or not the observations from more recent galaxy catalogs corroborate the assumption of homogeneity and isotropy at scales greater than 60 Mpc/h \citep{2009MNRAS.399L.128S, 2009MPLA...24.1743C},  or if instead the galaxies are grouped into highly structured hierarchical pattern, with properties of fractality \citep{citeulike:2502783, 2011ARep...55..324V}. 

In the last decade, the interest in the fractal analysis of the galaxy distribution has yielded promising results, becoming a fertile field for cosmological physics research \citep{2005astro.ph..5185B}. In particular, we do not know if the cold dark matter distribution have a fractal behaviour in the same way that is observed in the galaxy clustering. Some authors propose the same fractal dimension for the galaxy distribution and for the dark matter, where galaxies act as tracers of dark matter \citep{2008pc2..conf...60B}, while other researchers assign different fractal dimension; D=3 for dark matter, i.e. homogeneity in accord with cosmological principle and D=2 for galaxy clustering \citep{1998A&A...339L..85D}. Authors like \citet{0295-5075-71-2-332} suggest the use of multi-fractal models in order to analyse the dark matter haloes in numerical N--body simulations. He found that haloes of similar mass have a fractal distribution with a given dimension that grows as the mass diminishes. 

The homogeneous self-gravity matter can develop fractal structure \citep{1998CeMDA..72...91C}. For this reason the N--body cosmological simulations can be used like a numerical laboratory where the fractal analysis of matter under gravitational interaction can be performed. Besides, in simulations with sufficient number of particles it is possible to detect the fractal behaviour with the scale, including the transition to homogeneity. The statistical algorithms used in the analyses of the simulation data can be applied subsequently to galaxies surveys \citep{citeulike:2502783}. 

In this paper we present the calculations of mass-radius fractal dimension, mono-fractal lacunarity, multi-fractality and lacunarity spectrum of dark matter haloes distribution from the Millennium Simulation Project. We propose the use of multi-fractal dimension with the complementary measure of the lacunarity spectrum as a method for determining the homogeneity scale, i.e., the distance beyond which the transition to homogeneous regime should occur in a fractal matter clustering. 

The paper is divided as follows: Section 2 presents the mathematical foundations involved in the fractal approach to the clustering of large scale matter; section 3 summarises the characteristics of the Millennium simulation, focusing on the cosmological parameters and scale used as the basis of the simulation; section 4 discusses the calculation of mass-radius fractal dimension and lacunarity as functions of radial distance in the clustering of dark matter haloes from the Millennium Simulation; section 5 deals with the calculation of  the multi-fractal spectrum and the spectrum of lacunarity on the same data from the Millennium Simulation; in Section 6 the discussion of the fractal results is presented and finally the conclusions are given in section 7.

\section{Theoretical development}

\subsection{Dimension Concept}

The most general definition for the dimension measure was developed by Felix Hausdorff in 1918 \citep{springerlink:10.1007/BF01457179},  looking to define the dimension concept for any metric space. He first considered the number $N(\epsilon )$ of balls of radius $\epsilon$ needed to cover a subset $A$ completely. The  d-dimensional measure is proportional to the limit:

\begin{equation}
 h^{d}(A)= \lim_{\epsilon \to 0}{\mathit{}N(\epsilon)\epsilon^{d}}
\end{equation}

Hausdorff proved that there is only one value $d$ for which this is different from zero or infinity. For this value is satisfied that:
\begin{equation}
 h^{d}(A)=\left\{\begin{matrix}\infty & \mathrm{ if }\ d<D_{H}(A){}\\
                0\ & \mathrm{ if }\ d>D_{H}(A){}\end{matrix}\right.
\end{equation} where $D_H$ is by definition the Hausdorff dimension for the subset $A$,  with $D_H$  the value for which $N(\epsilon )\propto 1/\epsilon ^{d}$.  In order to decide whether a set is fractal or not it is necessary to compare the above definition with the topological dimension,  defined in a simple way as the number of independent directions in which one can move around a given point of the set. In this form Mandelbrot define a fractal as a set for which the Hausdorff dimension strictly exceeds its topological dimension. \citep{mandelbrot1983fractal}

It is not easy to determine the Hausdorff dimension for galaxy clustering, although there are methods developed to apply the Hausdorff dimension, like the Minimal Spanning Tree \citep{martinez2002statistics}. For this reason it is necessary to use alternative concepts of fractal dimension for the analysis of the matter distribution on a large scale in the universe. One of the most used concepts is the mass-radius fractal dimension. According to Blumenfeld and Mandelbrot  \citep{PhysRevE.56.112},  given a sphere of radius $r$ in a space with Euclidean dimension $d$, which encloses a self-similar fractal structure, the total mass measure $M(r)$ enclosed by the sphere take the form:

\begin{equation}\label{Mass-radius}
M(r) = F{r}^{D_{m}}
\end{equation} here $M(r)$ represents the number of galaxies (or in our case dark matter haloes)  in a sphere of radius $r$, $D_{m}$ is the fractal mass-radius dimension, and $F$ is a constant related to the average distance between nearest neighbours. The exponent $D_{m}$, the fractal dimension is smaller than the embedding Euclidean dimension $d$ for inhomogeneity condition in the set. 

Although the mass-radius dimension has the advantage of extending the measure of fractal dimension of the set on large scales, its accuracy is limited as it does not move the centre of observation, skewing the statistical analysis. Therefore it is necessary to refine the determination of fractality including average looking at several points in the sample as centres. According to \citet{citeulike:2502783} it is possible to average the mass-radius fractal dimension moving on different occupied points  of the galaxy distribution. The average mass-radius relation is a fractal measure which reflects the conditional cosmological principle: The cosmos appears statistically the same to all observers situated on a galaxy (point of a fractal) but not in a region of void \citep{mandelbrot1983fractal}. 

To fully describe a fractal it is necessary to specify how the pre-factor $F$ behaves. For this reason it is necessary to use a complementary fractal quantity, the lacunarity, which is a general measure for the analysis of spatial patterns.

\subsection{Lacunarity Definition}

The majority of the studies aiming to determine the fractality of large-scale structures in the universe begin with the calculation of the fractal dimension, but fractals with similar dimensions can occupy the space where they are embedded in different ways. Therefore, it is necessary to recognise that the fractal dimension is not enough to uniquely characterise fractal sets,  and for this reason \citet{PhysRevE.56.112} proposed a complementary fractal measure, the lacunarity. 

The lacunarity appears in the context of the mass-radius fractal dimension, although we can also extend this concept for other fractal dimension definitions like the correlation dimension \citep{2002SPIE.4847...86M}. Lacunarity high values suggest the presence of large empty regions inside the clustering and therefore an increased heterogeneity, while fractals with low values indicate approximation to homogeneity.

The starting point for calculating the lacunarity is the pre-factor $F$  from the mass-radius relation,  Eq. (\ref{Mass-radius}). There is not a unique definition to describe the properties of the factor $F$,  but Mandelbrot  proposed a description based in series of factors of variability \citep{PhysRevE.56.112}, 

\begin{equation}
S_{k}=\frac{C_{k}(D)}{[C_{1}(D)]^{k}},
\end{equation} where $C_k(D)$ is the $kth$ cumulant, a statistical measure of the correlation designed to go to zero whenever any one or more quantities under study become statistically independent. The cumulants are related with the moments, in particular $k_1 = \mu_1$ is the mean, $k_2$ is the variance, and $k_3= \bigl\langle(X - \mu_1)^3\bigr\rangle$ for a variable $X$ under study. The simplest of this variability factors, is the second order variability factor $S_{2}$ through which the lacunarity $\Phi$ is defined as: 

\begin{equation}
\Phi =\frac{\bigl\langle(F - \bigl\langle F \bigr\rangle)^{2}\bigr\rangle}{\ \bigl\langle F \bigr\rangle^{2}}=\frac{\bigl\langle F^{2}\bigr\rangle}{\bigl\langle F\bigr\rangle^{2}}-1.
\end{equation}

The Lacunarity analysis complements the studies of fractal sets enabling us to distinguish between similar fractal patterns which occupy the space in a different form. \citet{martinez2002statistics} applied the above definition of lacunarity to the Las Campanas survey data and the corresponding fractal Levy flight model. They found great difference in lacunarity behaviour with similar fractal mass-radius dimension. 

The average mass-radius dimension and the lacunarity are sufficient  to characterise sets where the fractal calculation of matter clustering at different scales and levels of mass density are the same. In the case of scale dependency,  we must go beyond in the characterising of the large scale matter distribution. This kind of analysis can be performed by the multi-fractal formalism.

\subsection{Generalised Dimension}

The dimension definitions presented above, represent particular cases of the multi-fractal spectrum of generalised dimension, method applied by the first time to the large scale structure in the universe by \citet{1992PhyA..185...45C}.  The large scale matter distribution can be characterised with a fractal spectrum defined from a generalised correlation. First, is necessary to introduce the correlation integral $C_{2}$, a function capable to measure the number of neighbours that on average a chosen centre has within a distance  $r$. We outline the development in accordance with the notation described in \citet{2008MNRAS.390..829B}:

\begin{equation}\label{C2}
 C_{2}(r)=\frac{1}{\mathit{NM}}\sum_{i=1}^M n_{i}(r),
\end{equation} where $N$ is the total number of particles inside the distribution, $M$ is the number of particles used as centres and the summation is performed over the set of chosen centres. $n_{i}(r)$ is the number of particles within a radial distance $r$ from a particle at the point $i$, defined as:
\begin{equation}\label{ni}
 n_{i}(r)=\sum_{j=1}^N \Theta(r-\mid \boldsymbol{x_i-x_j} \mid),
\end{equation}
where the summation is performed over all the particles in the sample. The coordinates of each particle in our space of three dimensions are denoted by $\boldsymbol{x_j}$, and $\Theta$ is the Heaviside function, defined such that $\Theta(x) = 0$ for $x < 0$ and $\Theta(x)=1$ for $x \geqslant 0$. The number of particles around each centre $n_{i}(r)$ is determined by counting the number of particles around that centre that lie inside a comoving sphere of radius $r$ from it.

From Equation (\ref {C2}) the correlation dimension is defined similarly to the definition of mass-radius dimension:
\begin{equation}\label{D2}
 C_{2}(r) \approx  r^{D_{2}}
\end{equation}

So that the correlation dimension is calculated as a derivative:

\begin{equation}
  D_{2}=\frac{d\log C_{2}(r)}{d\log r}
\end{equation}

From the correlation integral $C_{2}(r)$, Equation (\ref{C2}), the generalised correlation integral can be defined as:

 \begin{equation}
 C_{q}(r)= \frac{1}{\mathit{NM}}\sum_{i=1}^M [n_{i}(r)]^{q-1} 
\end{equation} 
where $M$ is the number of centres, $N$ is the total number of particles included in the sample, $n_{i}(r)$ is the same expression defined in the Equation (\ref {ni}) and $q$ is called the structure parameter, which corresponds to an arbitrary real number. From this generalised correlation it is possible to do an expansion in powers of $log(r)$ as described by \citet{1997Chaos...7...82P} and to calculate directly the multi-fractal dimension and the lacunarity spectrum:

 \begin{equation}
log \left[ C_{q}(r)^{1/(q-1)}\right] =D_{q}log(r)+log(F_{q})+O \left( \frac{1}{log(r)}\right).
\end{equation}
Keeping only the first two terms on the right side, we have the relation between the generalised correlation integral and the generalised fractal dimension:

\begin{equation}
C_{q}(r)^{1/(q-1)}=F_{q} r^ {D_{q}}
\end{equation}

In this manner the generalised dimension and the generalised lacunarity can be defined in the same way as for the mass-radius fractal dimension, the generalised dimension $D_{q}$:

\begin{equation}
  D_{q}=\frac {1}{(q-1)} \frac{d\log C_{q}(r)}{d\log r},
\end{equation} and the corresponding generalised lacunarity from the pre-factor  $F_q$ for every structure parameter $q$:

\begin{equation}
\Phi _{q}=\frac{\bigl\langle(F_{q} - \bigl\langle F_{q} \bigr\rangle)^{2}\bigr\rangle}{\ \bigl\langle F_{q} \bigr\rangle^{2}}=\frac{\bigl\langle F_{q}^{2}\bigr\rangle}{\bigl\langle F_{q} \bigr\rangle^{2}}-1.
\end{equation}

If for any $q_1\neq q_2$,  $D_{q_1} = D_{q_2}$ is verified, it is said that the distribution is a homogeneous fractal (mono-fractal). For $q \geqslant 1$, $D_q$  explores the scaling behaviour  in high density environments (clusters and superclusters) and  for $q < 1$ values, $D_q$ explores the scaling behaviour in low-density environments, i.e., voids \citep{2009MNRAS.399L.128S}. In the event that the distribution of dark matter haloes undergo the transition to homogeneity, all values of  the fractal dimension must close to the physical space dimension $(D_q \to 3) $ and the spectrum of lacunarity must tend to zero $(\Phi _{q} \to 0)$, at the same radial distance $r$. 

Next we will apply the concepts of average mass-radius dimension,  generalised dimension and generalised lacunarity to the dark matter haloes spatial distribution generated by the Millennium dark matter simulation.
 
\section{Millennium Simulation}

The Millennium Simulation \citep{2005Natur.435..629S} is one of the most important computational efforts in contemporary cosmology. It uses the most accepted parameters inside the standard cosmology, with particles evolving since redshift $z=127$ until the present $z=0$. Although the standard cosmological model contains two ingredients which have not yet been verified by laboratory experiments (the components of dark matter and dark energy), the $\Lambda$CDM model is almost universally accepted by cosmologists as the best description of the present observational data. The spatial geometry is very close to flat and the initial perturbations which should have given origin to galaxies are Gaussian, adiabatic, and nearly scale-invariant. The cosmological parameters assumed inside this computational algorithm are summarised in Table \ref {table: Millparameters}.

\begin{table}
\centering
\begin{tabular}{llllll}
 \hline
$\Omega_m$ & $\Omega_b$ & $\Omega_\Lambda$ & $h$ & $\sigma_8$ & $n$ \\
\hline
$0.25$ & $0.045$ & $0.75$ & $0.73$ & $0.9$ & $1$
\end{tabular}
\caption{Cosmological parameters for the Millennium Simulation}
\label{table: Millparameters}
\end{table}

Here the density parameters $\Omega_i = \rho_i /  \rho_{crit}$ are defined from the critical density $ \rho_{crit}=3H_0^2/8\pi G $. The total present matter density composed by dark matter and baryonic matter $\Omega_m= \Omega_b+\Omega_{dm}$, the dark energy density $\Omega_\Lambda$ with $\Omega_m+ \Omega_\Lambda=1$, the Hubble parameter $h=H_0/(100km s^{-1}Mpc^{-1})$, $\sigma_8$ the linear-theory power spectrum variance in spheres of radius 8 Mpc/h and the spectral index $n$ used to describe the density perturbations in Fourier space, where $n=1$ corresponds to Harrisson-Zeldovich scale invariant spectrum. 

The simulation is implemented with the length of the side of the simulation cube $L_{box} = 500 Mpc/h$, the number of particles $N_{part}=2160^3\approx  10^{10}$ , with the resulting particle mass $m_p=8.6 \times 10^8 M_{\odot}/h$  and the gravitational softening length $\eta=5 kpc/h$. The Virgo consortium, provides free access to the database, which authorised users can entry using the Structured Query Language (SQL). It is expected that different researchers employ the information to explore how the structures of dark matter and galaxies evolve in the standard cosmology.  

We use the information provided by this simulation in order to perform a fractal analysis of the dark matter haloes clustering, aiming to find the way in which the dark matter haloes are grouped as a function of  the radial distance and to detect the scale of transition to homogeneity. 

\section{Fractal Mass-Radius Dimension and Lacunarity of Millennium Simulation data}

\begin{figure}
\includegraphics[width=1 \linewidth,clip]{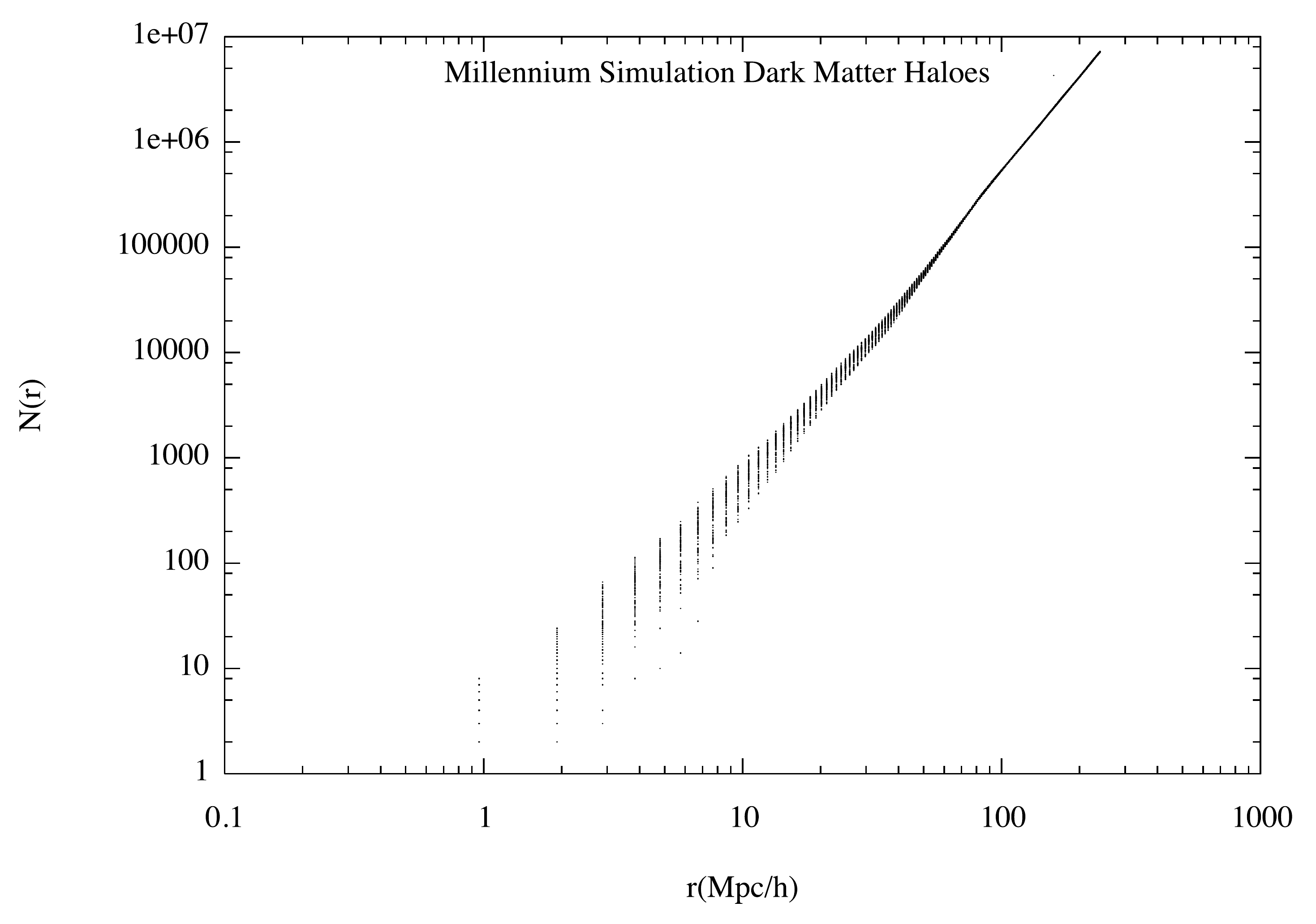}
 \caption{Millennium Simulation Fractal Analysis:  Number of dark matter 
 haloes \textit{vs} Radial distance  over 100 centres randomly chosen, in logarithmic scale. The initial data dispersion and the slope growth is shown. }
	\label{fig: lognlogr}
\end{figure}

The first step in fractal analysis of dark matter haloes clustering, for $z = 0$ from the millennium simulation, is the choice of a random sample of centres in the entire simulation ($\approx 1.5 \times 10^7$ dark matter haloes).  We chose randomly one hundred centres with the purpose to determine the functional relationship between the logarithm of the number of  dark matter haloes and the logarithm of the radial distance. We want to know if there is a linear function that leads to the determination of the fractal correlation dimension $D_m$ and the pre-factor $F$ as shown in Figure \ref{fig: lognlogr}.

In the case of a mono-fractal set, the expected relationship between the logarithm of the number of particles and the logarithm of radial distance should be close to a linear behaviour. By observing Figure \ref{fig: lognlogr} a large scatter in the relationship for distances $r < 20$ Mpc/h is evident. Also, it would be inappropriate to assign a single slope for the entire curve when a gradual increase of slope with the radial distance is observed, suggesting a fractal dimension growing with scale. It is therefore necessary to calculate the fractal dimension and the pre-factor $F$ in small sections of the curve for each centre, in order to determine if there is a radial distance dependency of the fractal quantities. 

\begin{figure*}
	\includegraphics[width=0.45\linewidth,clip]{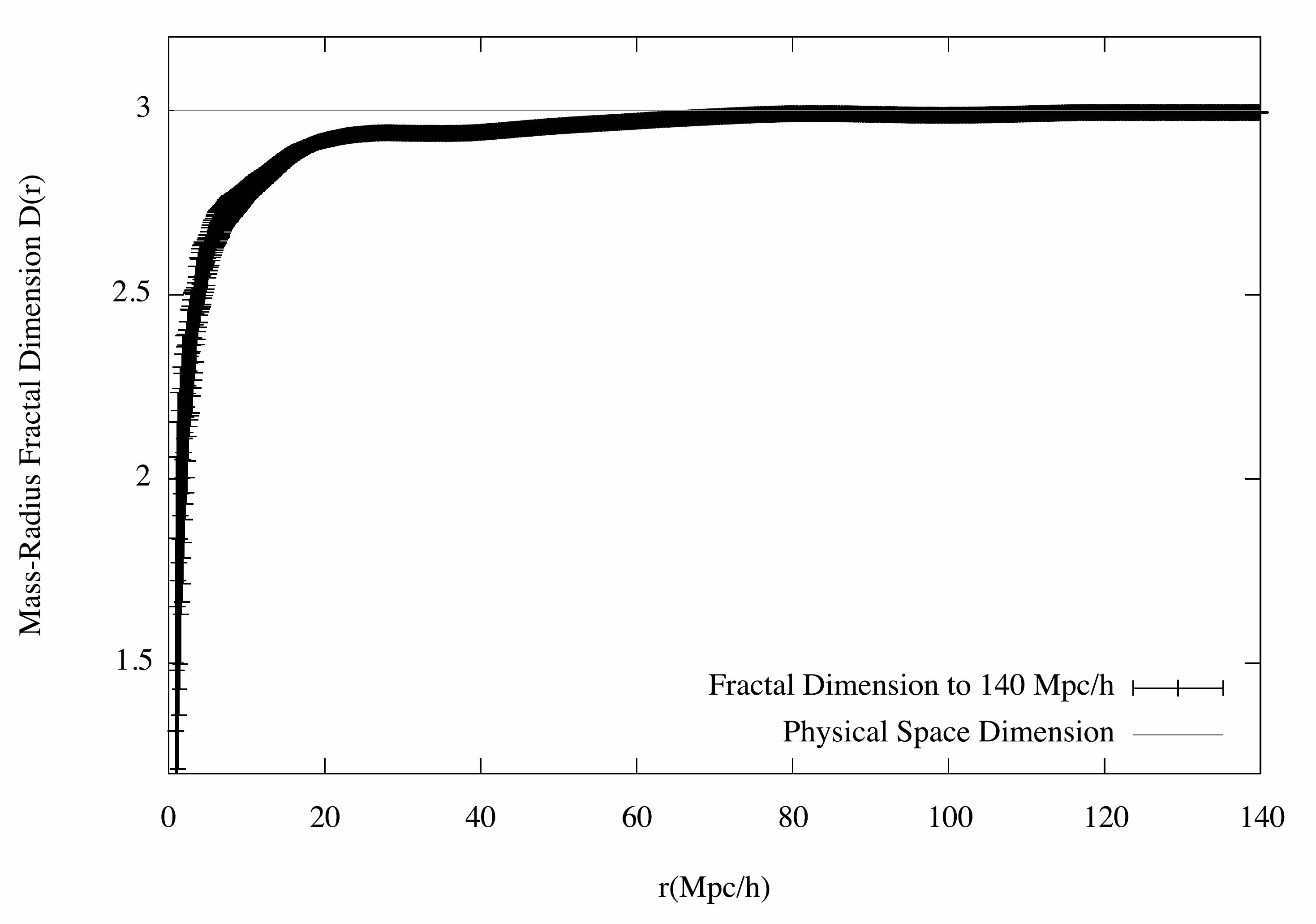}
	\includegraphics[width=0.45\linewidth,clip]{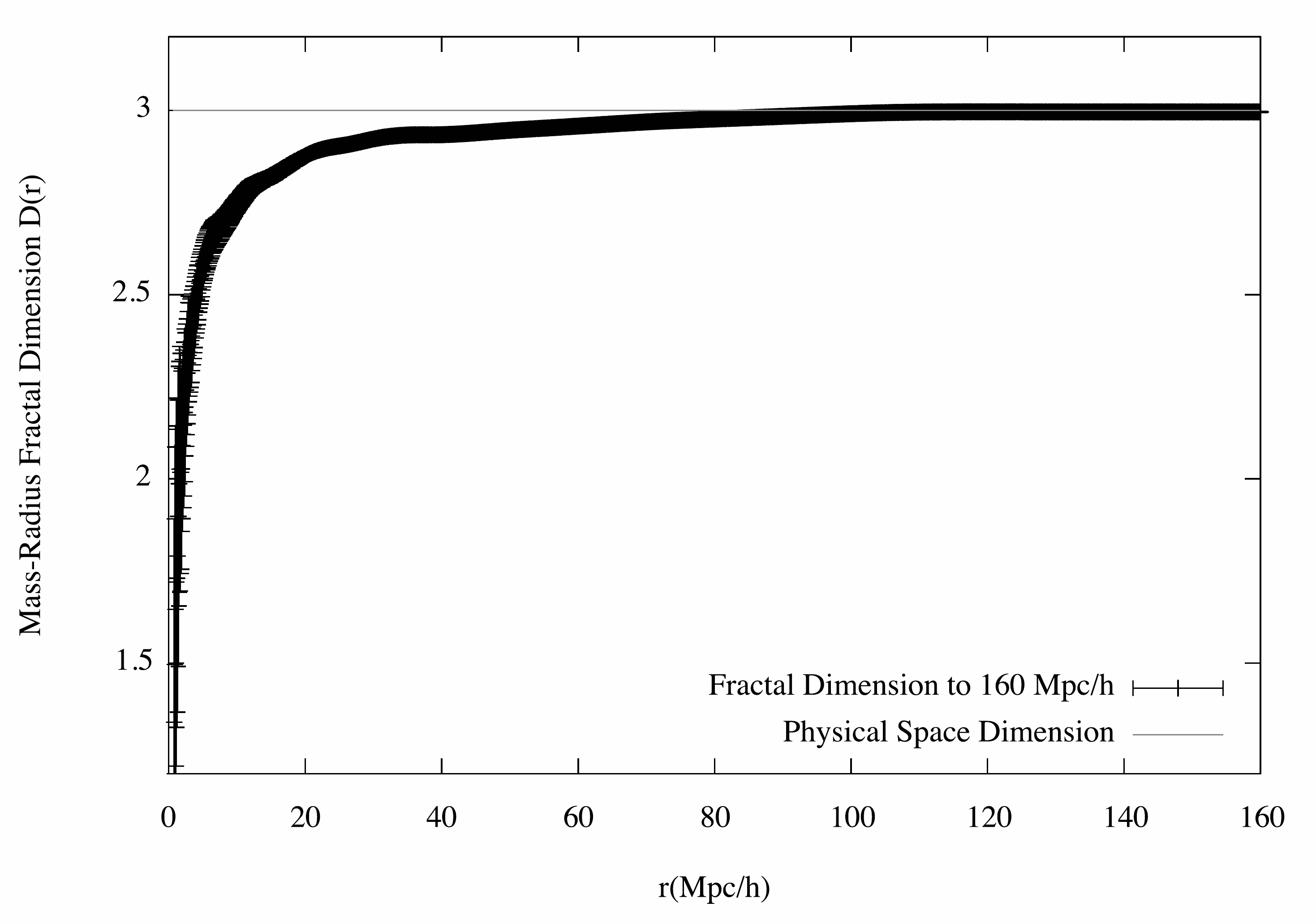}
         \includegraphics[width=0.45\linewidth,clip]{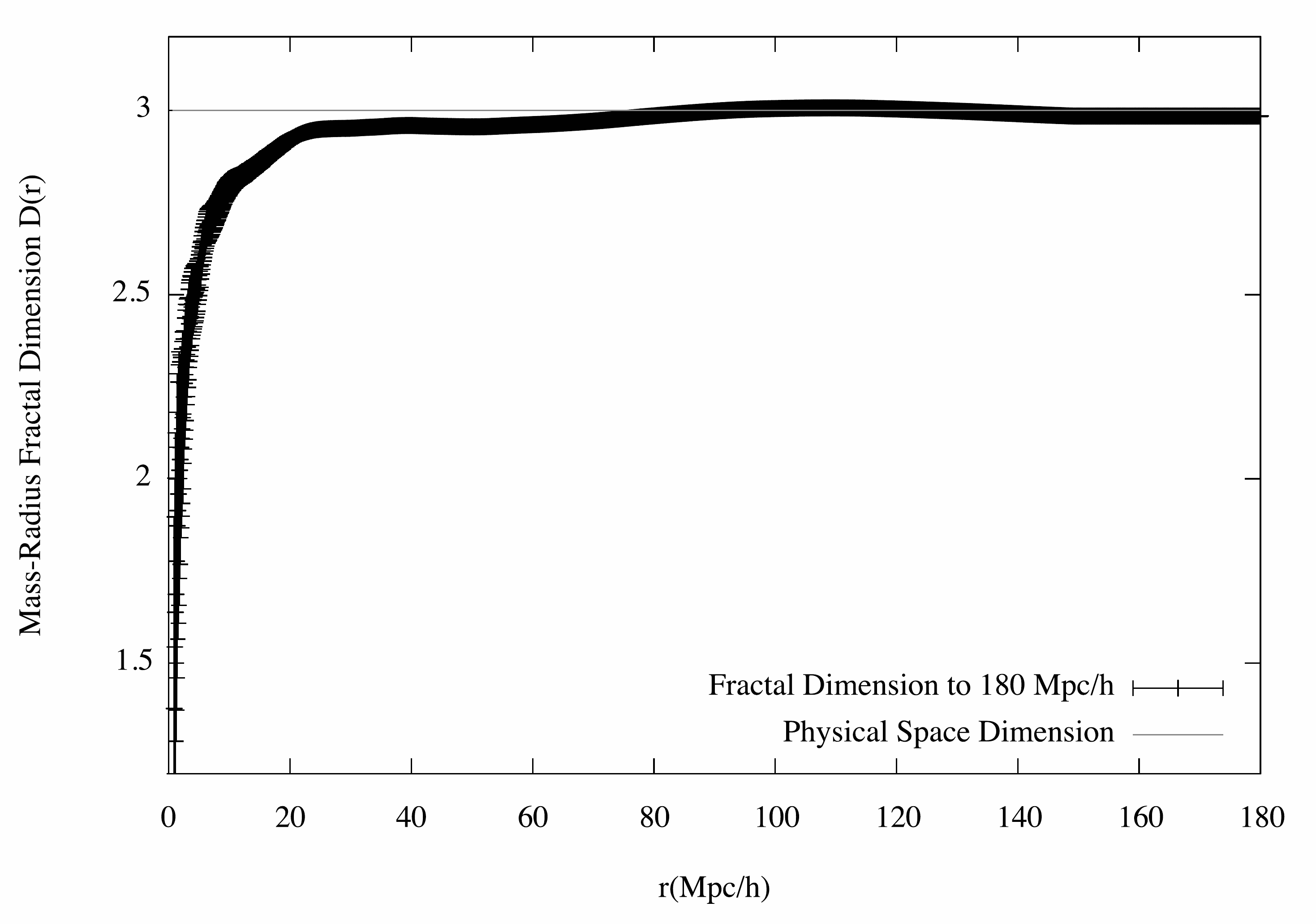}
         \includegraphics[width=0.45\linewidth,clip]{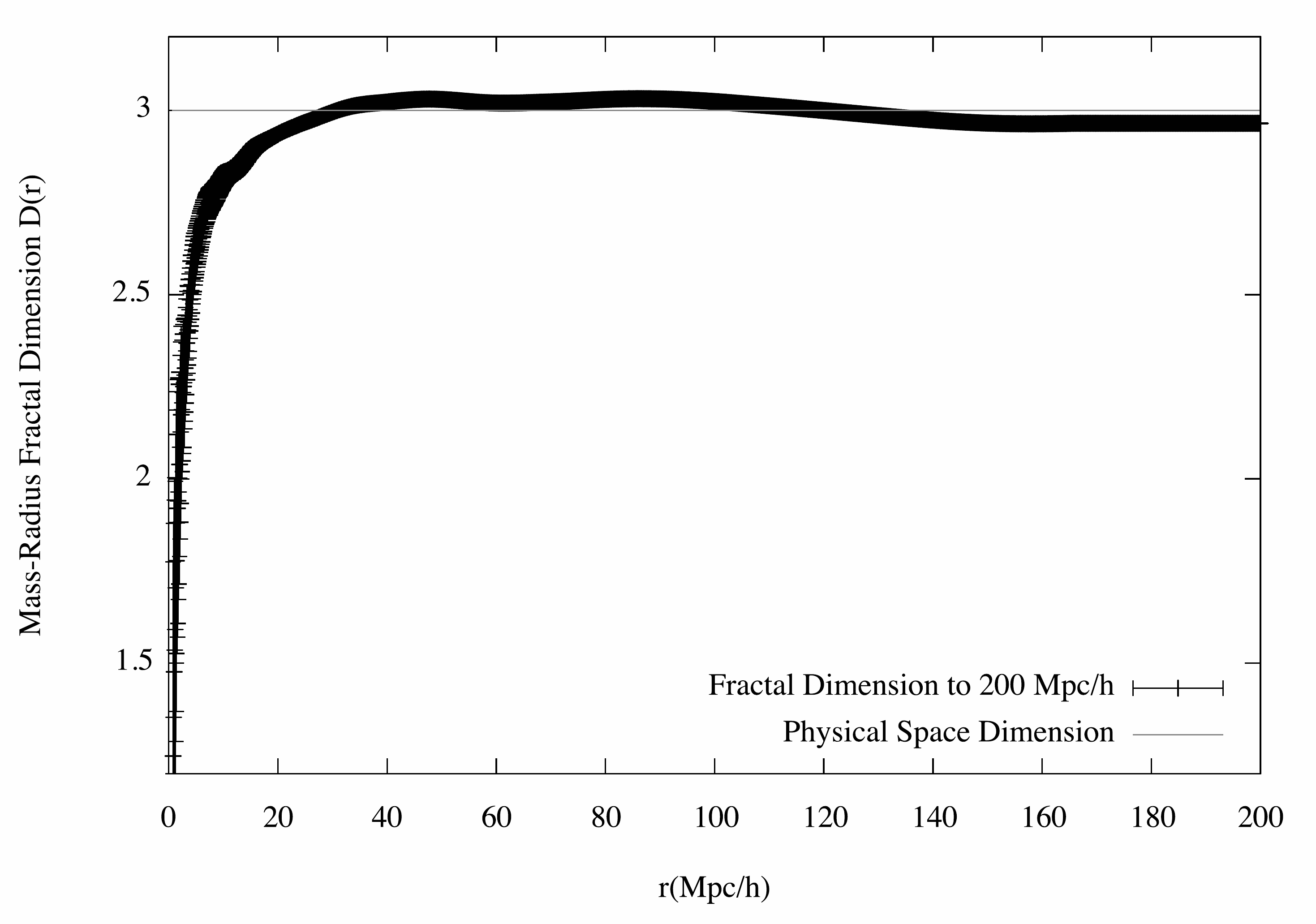}
        
         \caption{The average of mass-radius fractal dimension \textit{vs} radial distance for the dark matter haloes distribution in the Millennium simulation. Spurious homogenisation effects in the curves around 20 Mpc/h are observed to fractal calculations for depths larger than 180 Mpc/h.  $\pm1\sigma$ error bars are shown.}
	\label{dmvsr}
\end{figure*}

Dividing the curve in segments using the sliding window technique  \citep{ 2002SPIE.4847...86M, 2005PhRvE..72a6707R} and making a linear least-square fit through each set of  successive points in the log-log plot, it is possible to calculate the slope and intercept of the line that approximates the curve section over every point for each of the curves generated from each centre, and then determine the average over a thousand centres (more representative sample) with their respective standard deviation. We also analyse the behaviour of this average mass-radius fractal dimension as the sample size approaches to the limits of the simulation, as shown in Figure \ref{dmvsr}.

In Figure \ref{dmvsr}, the graphs for different depths have a similar behaviour; a rapid increase in the fractal dimension to radial distances about 20 Mpc/h followed by slower growth until it reaches the physical dimension of space to depths beyond the 100 Mpc/h. Those graphics with radial distances from the centres exceeding 180 Mpc/h, where the dark matter haloes are close to the edges of the simulation,  show that the average mass-radius fractal dimension at short distances is larger than the physical dimension of the space where the whole fractal is embedded, a physically unacceptable situation.  

\begin{figure*}
	\includegraphics[width=0.45\linewidth,clip]{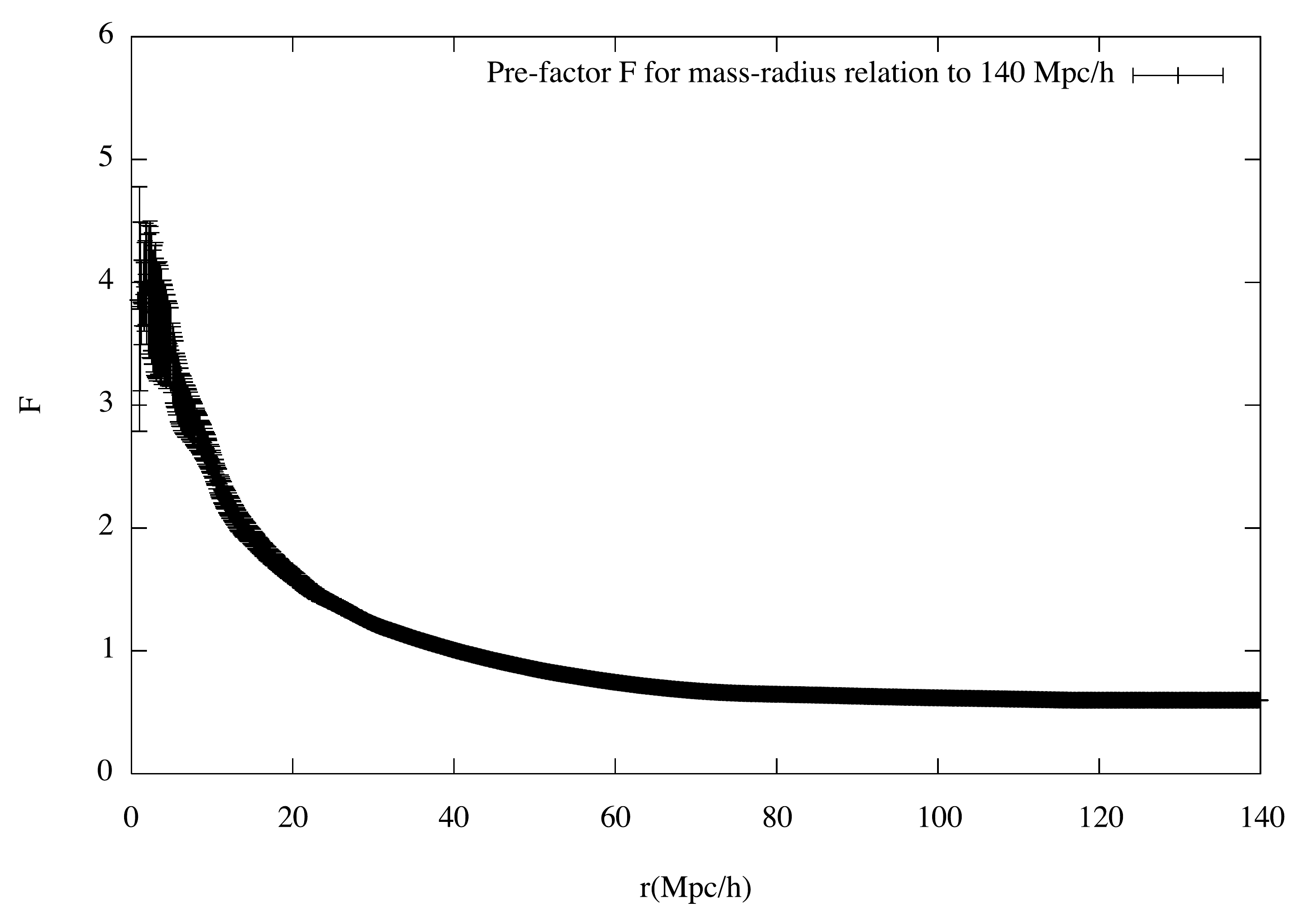}
	\includegraphics[width=0.45\linewidth,clip]{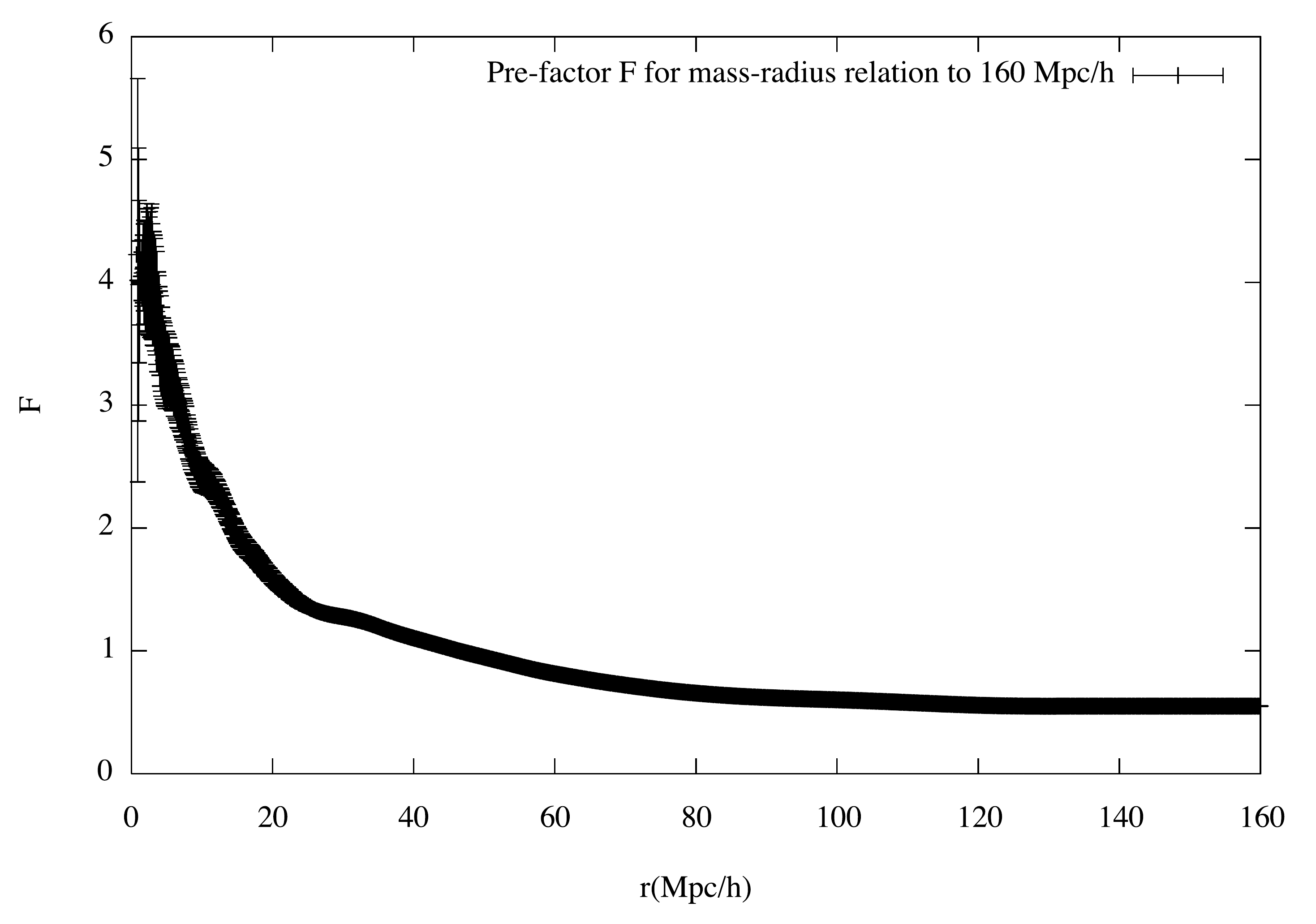}
         \includegraphics[width=0.45\linewidth,clip]{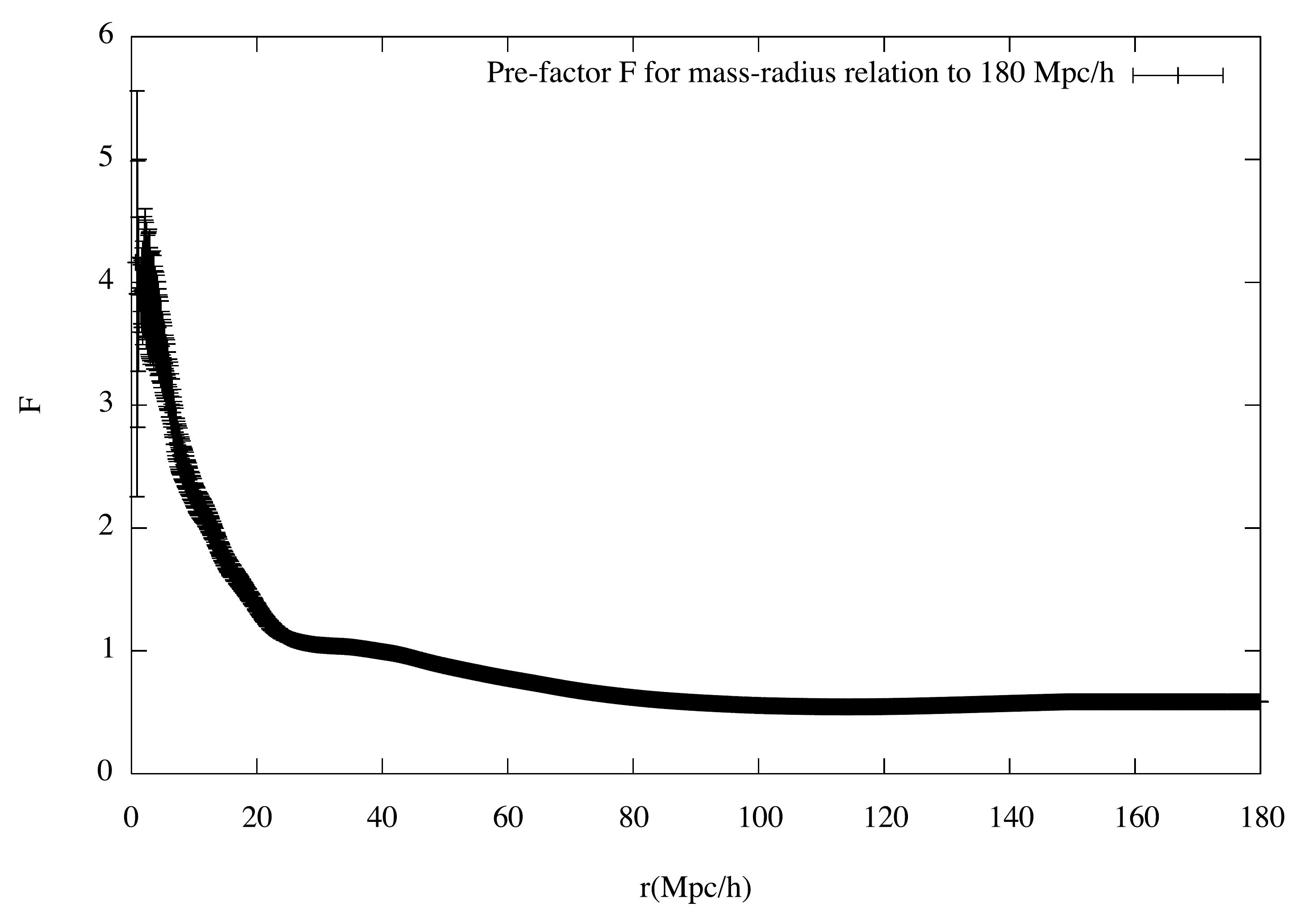}
         \includegraphics[width=0.45\linewidth,clip]{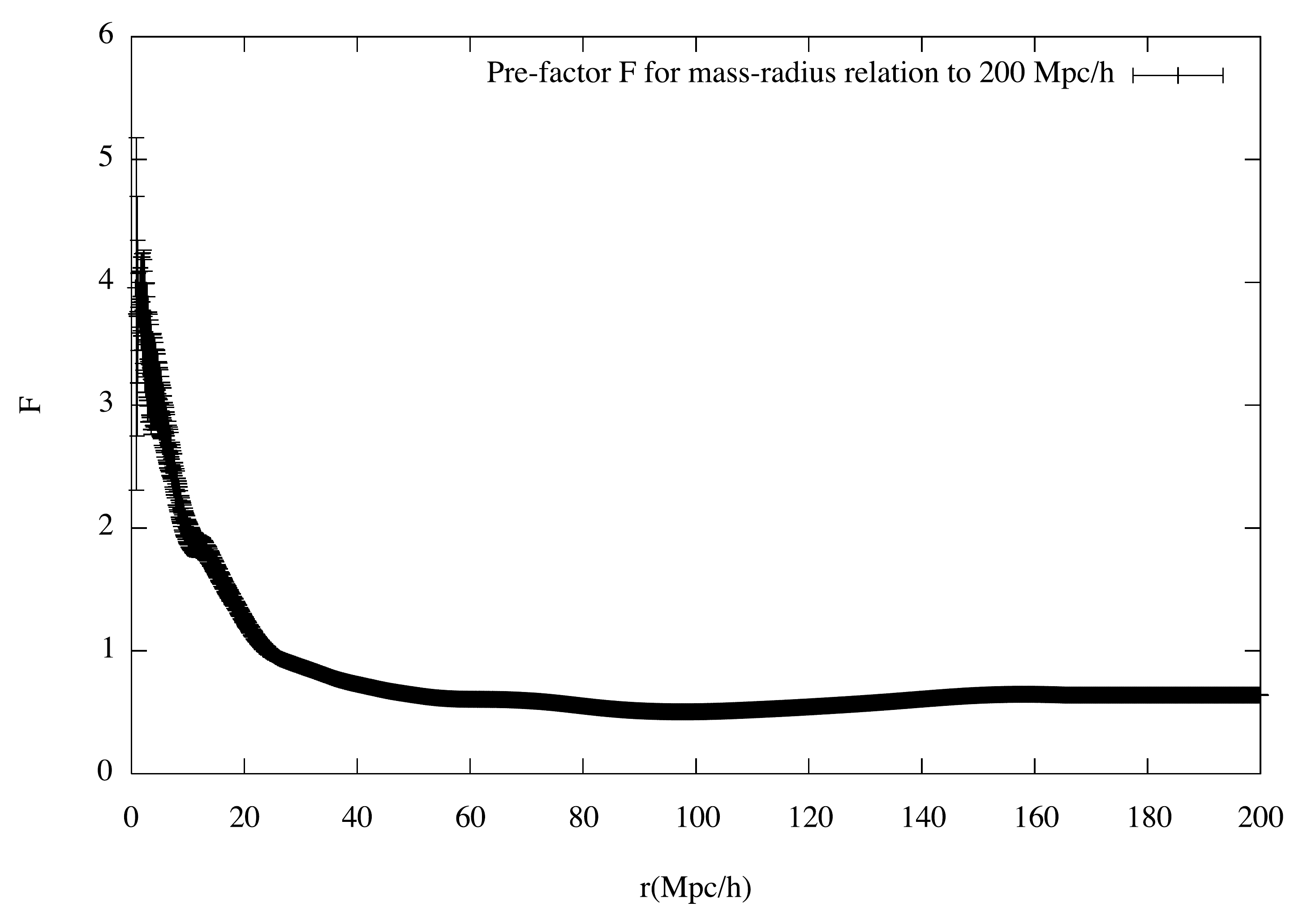}
        
         \caption{Fractal Pre-factor $F$ \textit{vs} radial distance for the dark matter haloes distribution in the Millennium simulation. There is a tendency to achieve a constant value with increasing radial distance. $\pm1\sigma$ error bars are shown.}
	\label{fvsr}
\end{figure*}

The behaviour of the pre-factor $F$ is shown in Figure \ref{fvsr}. The values of the pre-factor change with scale, starting with high values near the centres and diminishing to reach a low constant value at distances larger than 100 Mpc/h. Therefore, the average distance between dark matter neighbours decreases with radial distance reaching a constant value for radial distances larger than 100 Mpc/h. A change at short distances of the behaviour of the curves for depths from the centres larger than 180 Mpc/h is detected.

This results confirm that there are spurious homogenisation effects in the calculations of the fractal dimension. Thus, it is necessary to calculate the fractal dimension over radial distances smaller than 180 Mpc/h, i.e., radial distances where the fractal dimension does not exceed $3\pm1\sigma$, as recommended by \citet{1989A&A...219....1W}. In our case the maximum radial distance that satisfies the above requirement is 160 Mpc/h.

\begin{figure*}
	\includegraphics[width=0.45\linewidth,clip]{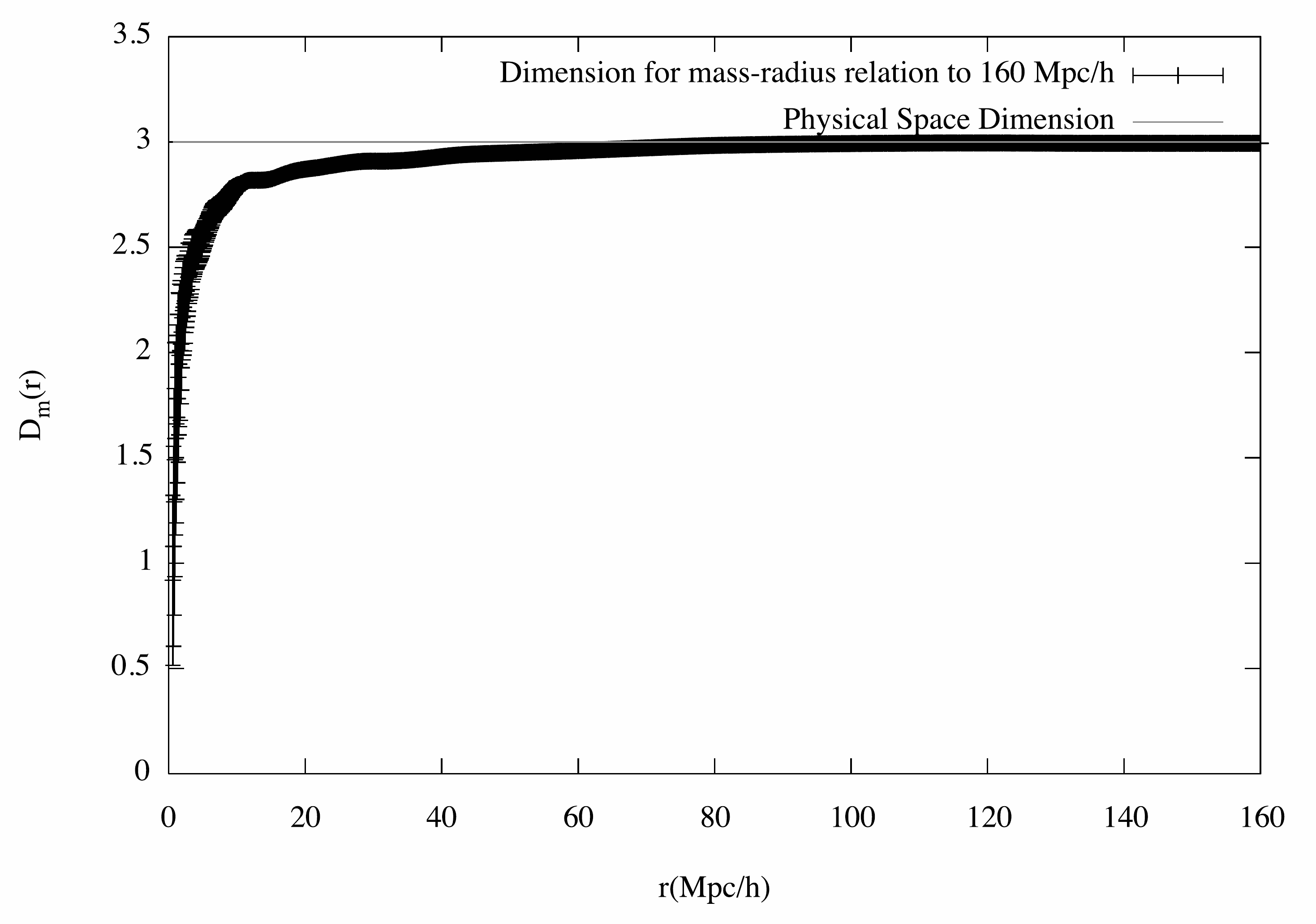}
	\includegraphics[width=0.45\linewidth,clip]{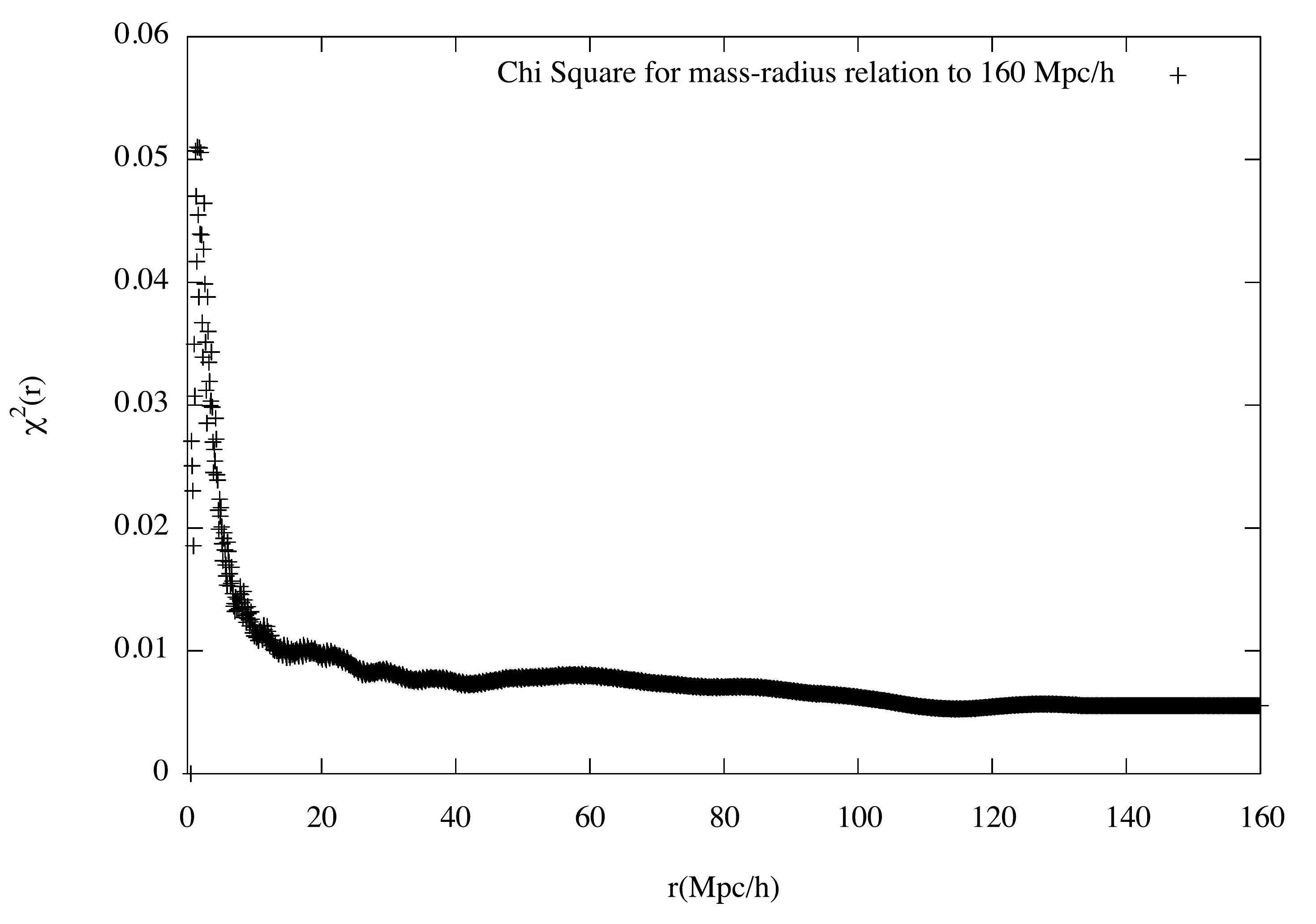}
	\includegraphics[width=0.45\linewidth,clip]{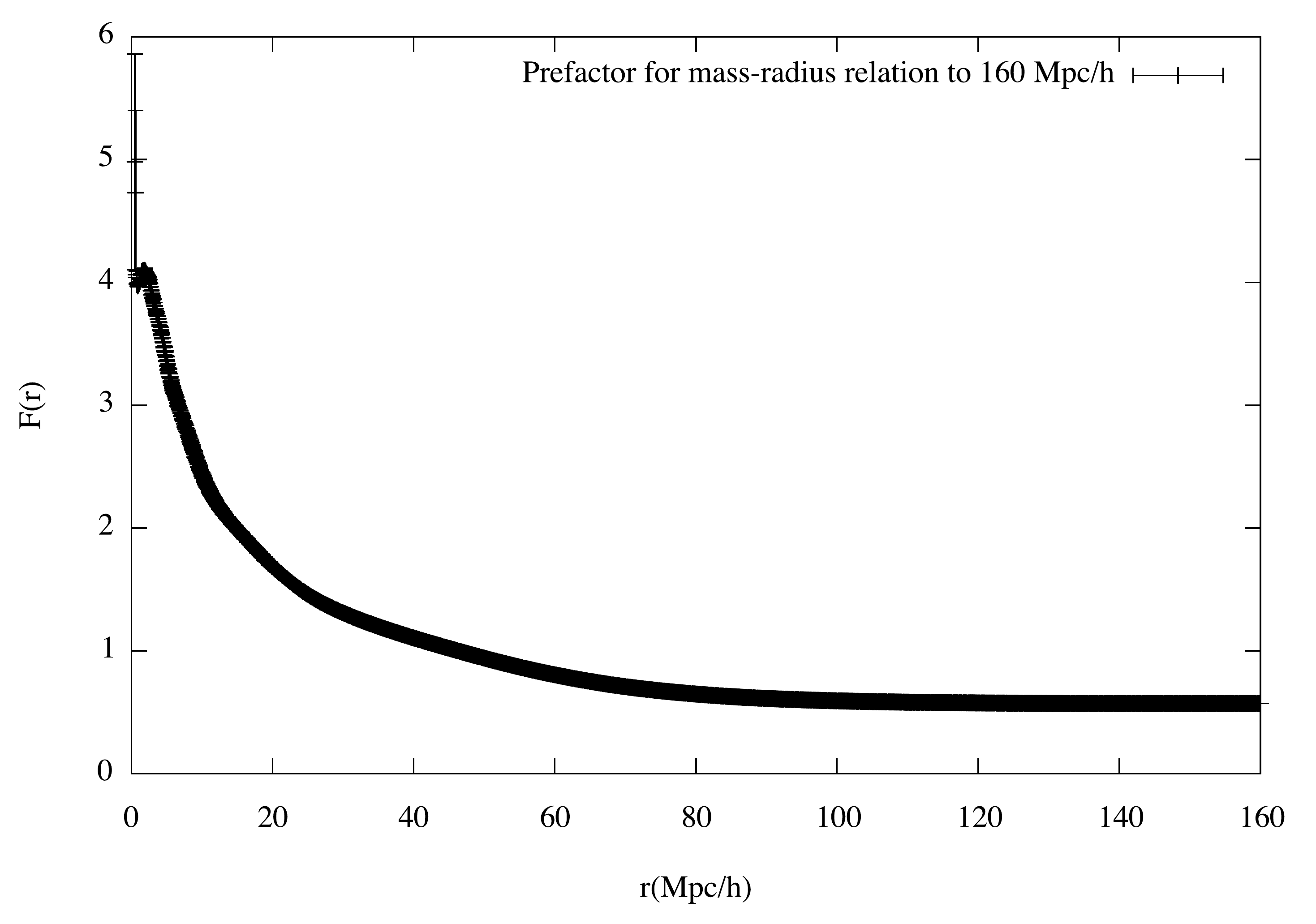}
	\includegraphics[width=0.45\linewidth,clip]{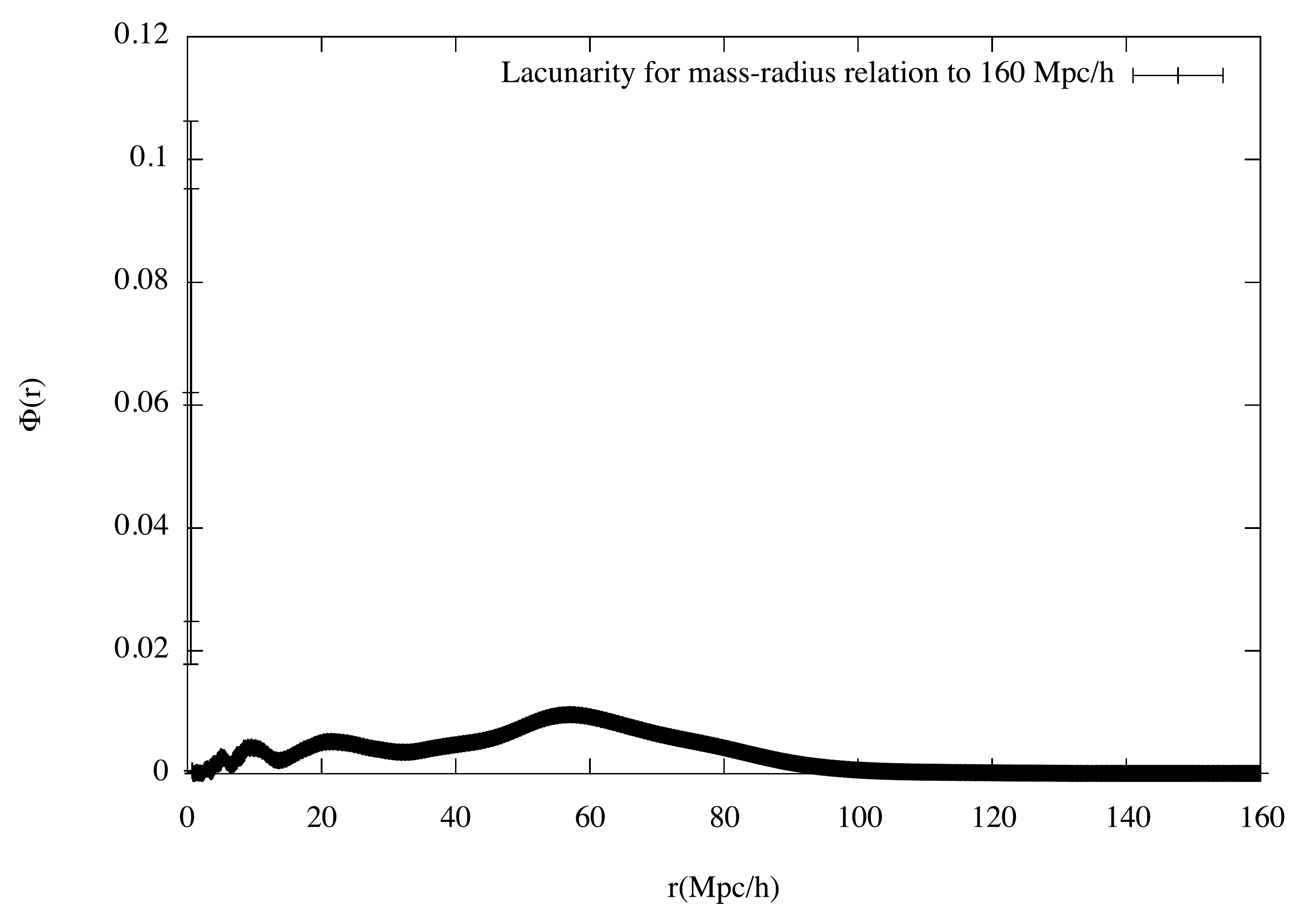}
	
 \caption{Mass-radius quantities calculated with the sliding window technique: The average of the mass-radius dimension $D_m$, the pre-factor $F$, the lacunarity $\Phi$ and the $\chi^2$ test \textit{vs} the radial distance, for the dark matter haloes distribution. $\pm1\sigma$ error bars are shown. The average $<\chi^2> = 7.9\times10^{-3}$, with $\chi^2 _{max}=5.1 \times10^{-2}$.}
	\label{resume160}
\end{figure*}

To complement the vision of the dark matter haloes fractal distribution we calculate the lacunarity up to 160 Mpc/h radial distances, taking into account the pre-factor of the sliding window least-square fit analysis to the same depth within the simulation. The lacunarity of the dark matter haloes distribution, Figure \ref{resume160}, shows a maximum value near each chosen centre, followed by oscillatory lacunarity behaviour at radial distances smaller than 60 Mpc/h. In addition, above 60 Mpc/h we see a smoothly decreasing lacunarity function with tendency to homogeneity at radial distances $\approx 120$ Mpc/h where lacunarity values approach to zero.

The fractal quantities show a strong reliance on the scale. The decrease in the lacunarity curve and the scale dependence of the average mass-radius fractal dimension with the radial distance suggest us that the clustering of dark matter haloes do not behave like a mono-fractal. Therefore it is necessary to study the dark matter haloes distribution from the multi-fractal viewpoint. 

\section{Multi-fractality and Multi-lacunarity of Millennium Simulation data}

\begin{figure*}
	\includegraphics[width=0.45\linewidth,clip]{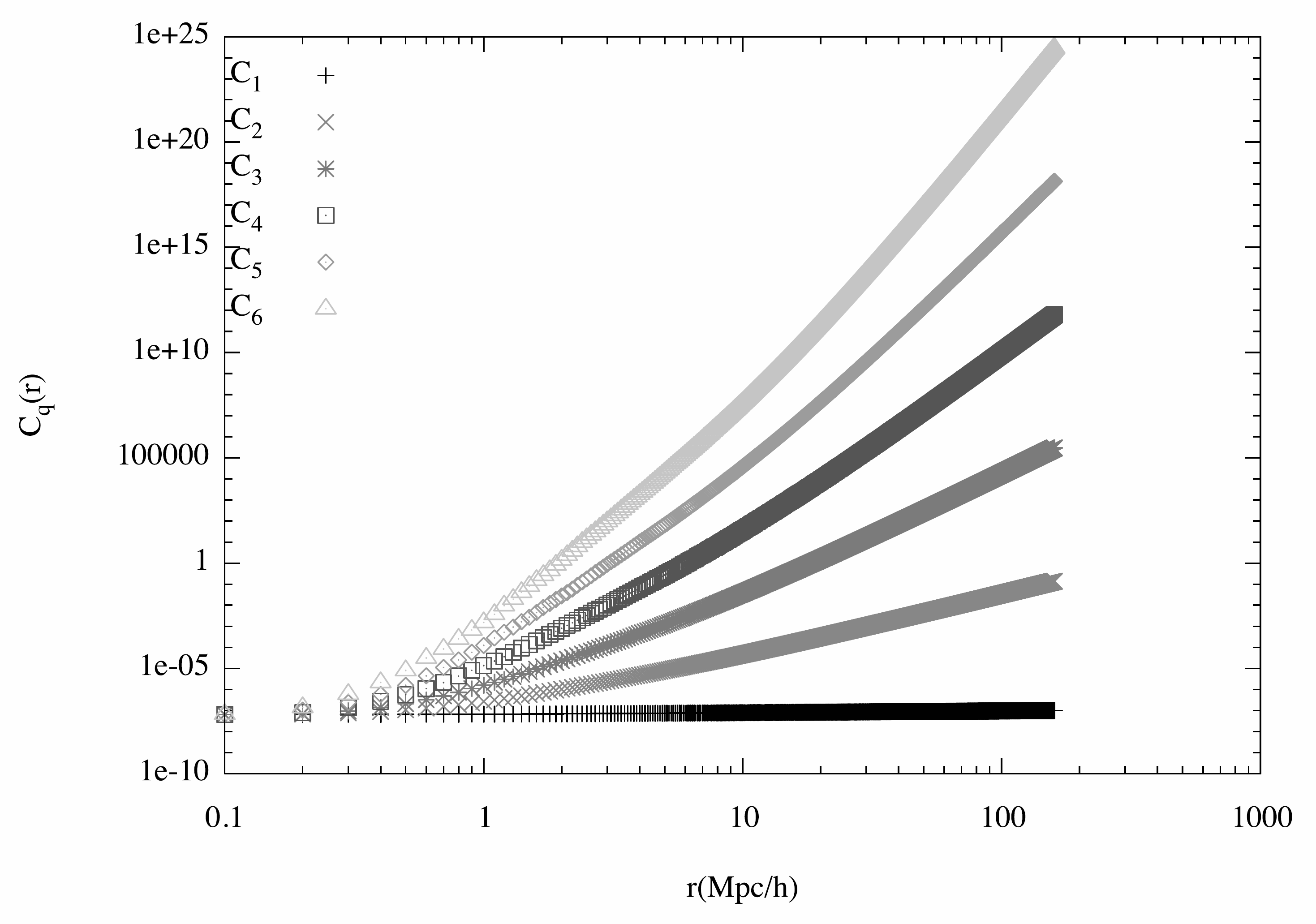}
	\includegraphics[width=0.45\linewidth,clip]{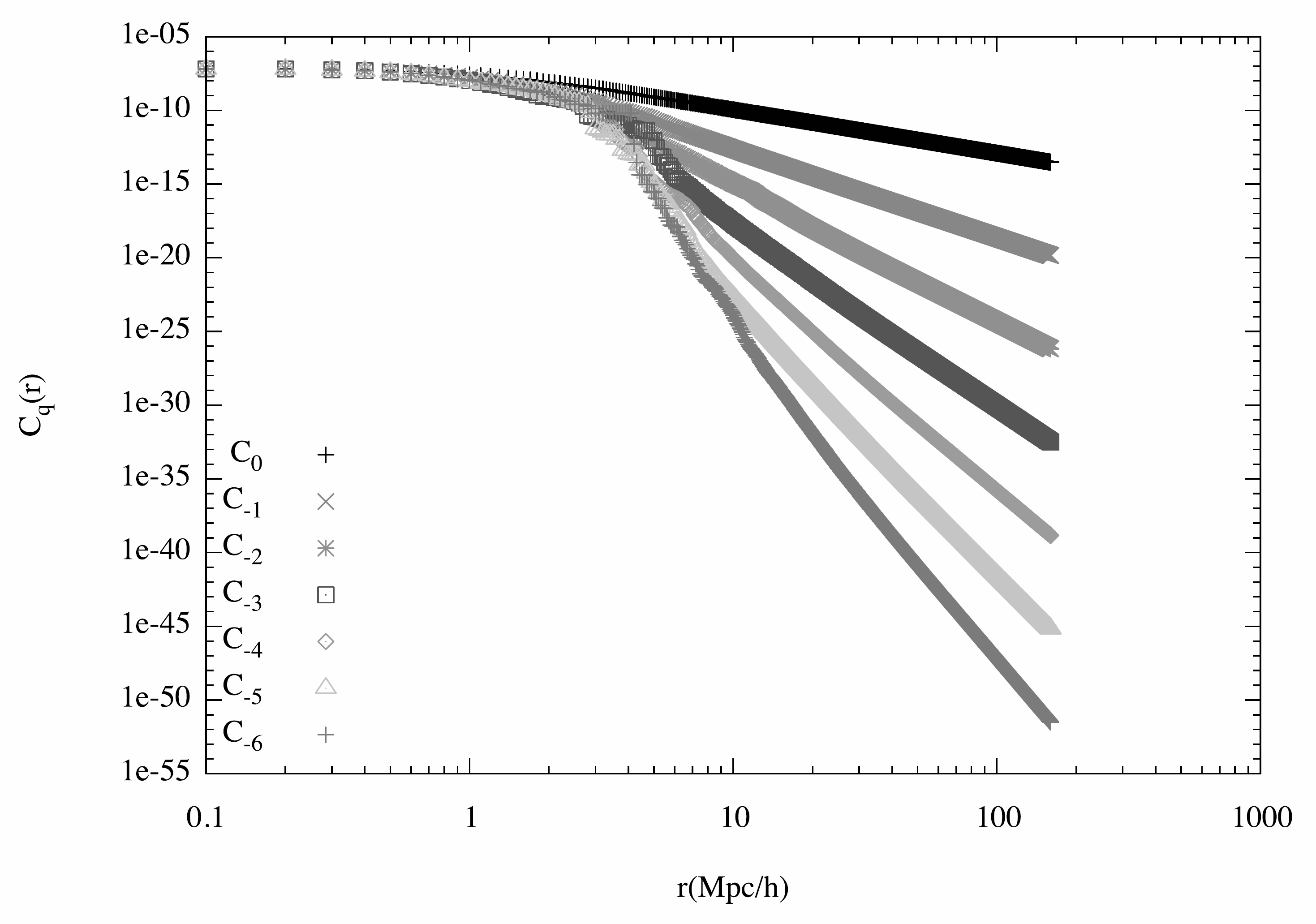}

 \caption{Generalised Correlation Integral $C_q(r)$ for all the values of $q$ studied in this paper for the Millennium simulation dark matter haloes. Over densities  $q \geqslant 1 $ left, low densities $q < 1$ right. All graphics in log-log scale}
	\label{Cqr}
\end{figure*}

In the last section we have shown that the Millennium Simulation dark matter haloes average mass-radius fractal dimension have a radial distance dependency with a lacunarity descending to zero around 120 Mpc/h. Now, the results of the generalised correlation integral for twelve different values of $q$ are presented in Figure \ref{Cqr}. This calculations include overdense regions $(q \geqslant 1)$ and low density regions $(q < 1)$. For $q =1$ the numerical limit of the generalised correlation integral is shown. To avoid the problems of spurious homogenisation found in the mass fractal dimension,  the multi-fractal calculus is limited to depths still far from the borders of the simulation. Based on the average mass-radius fractal behaviour, the maximum depth for the calculations in the Millennium data is 160 Mpc/h. 

\begin{figure*}
	\includegraphics[width=0.45\linewidth,clip]{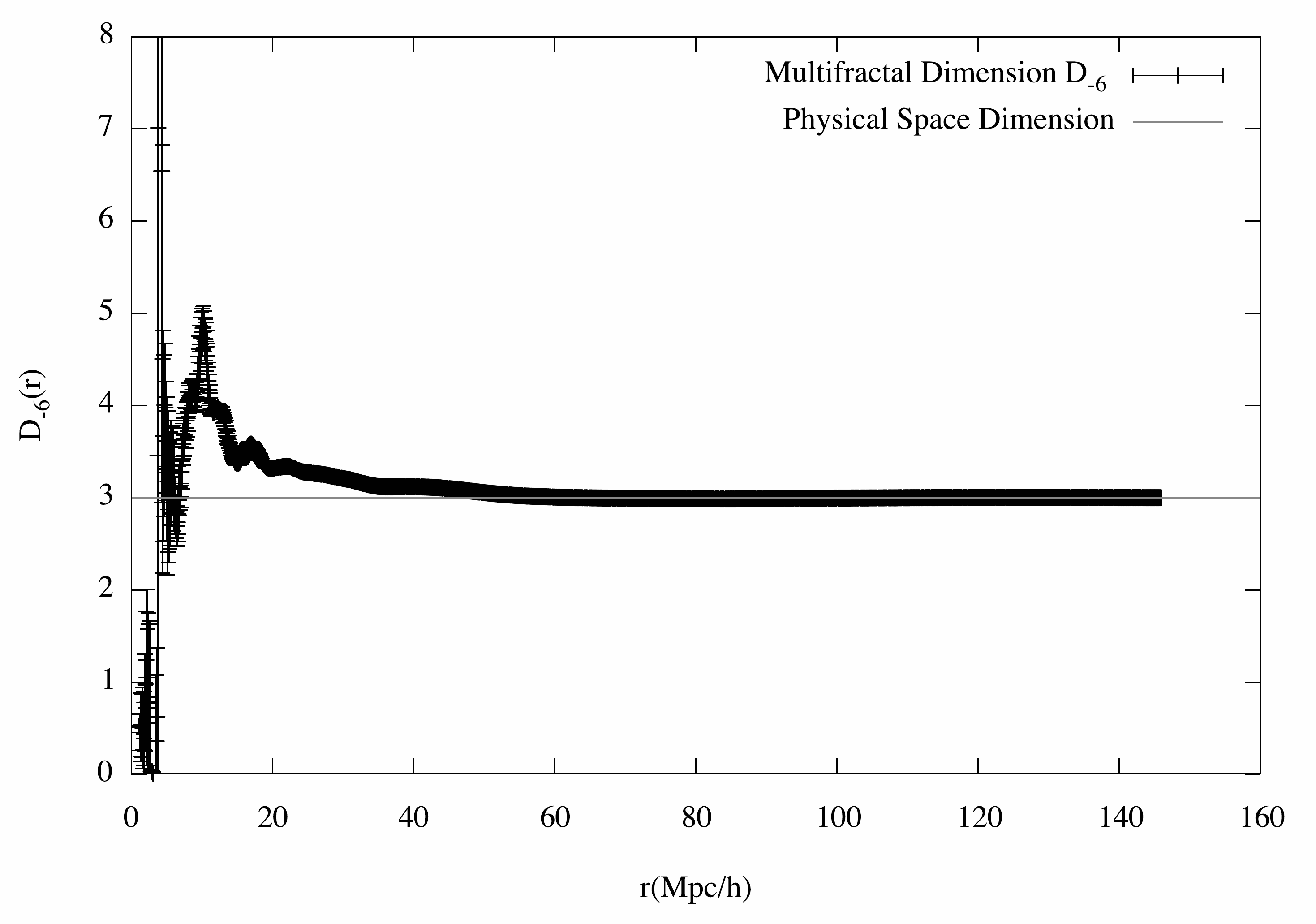}
	\includegraphics[width=0.45\linewidth,clip]{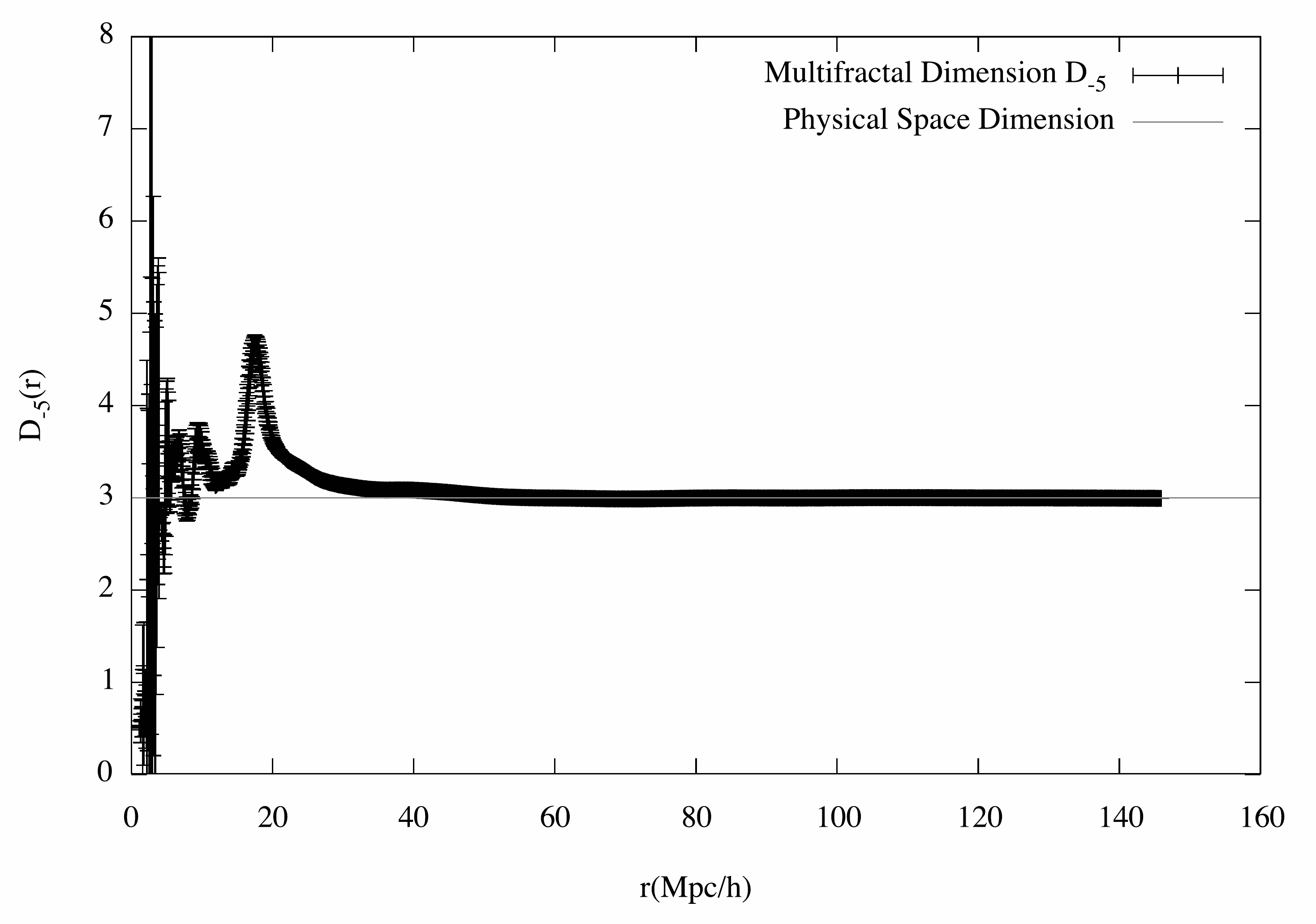}
	\includegraphics[width=0.45\linewidth,clip]{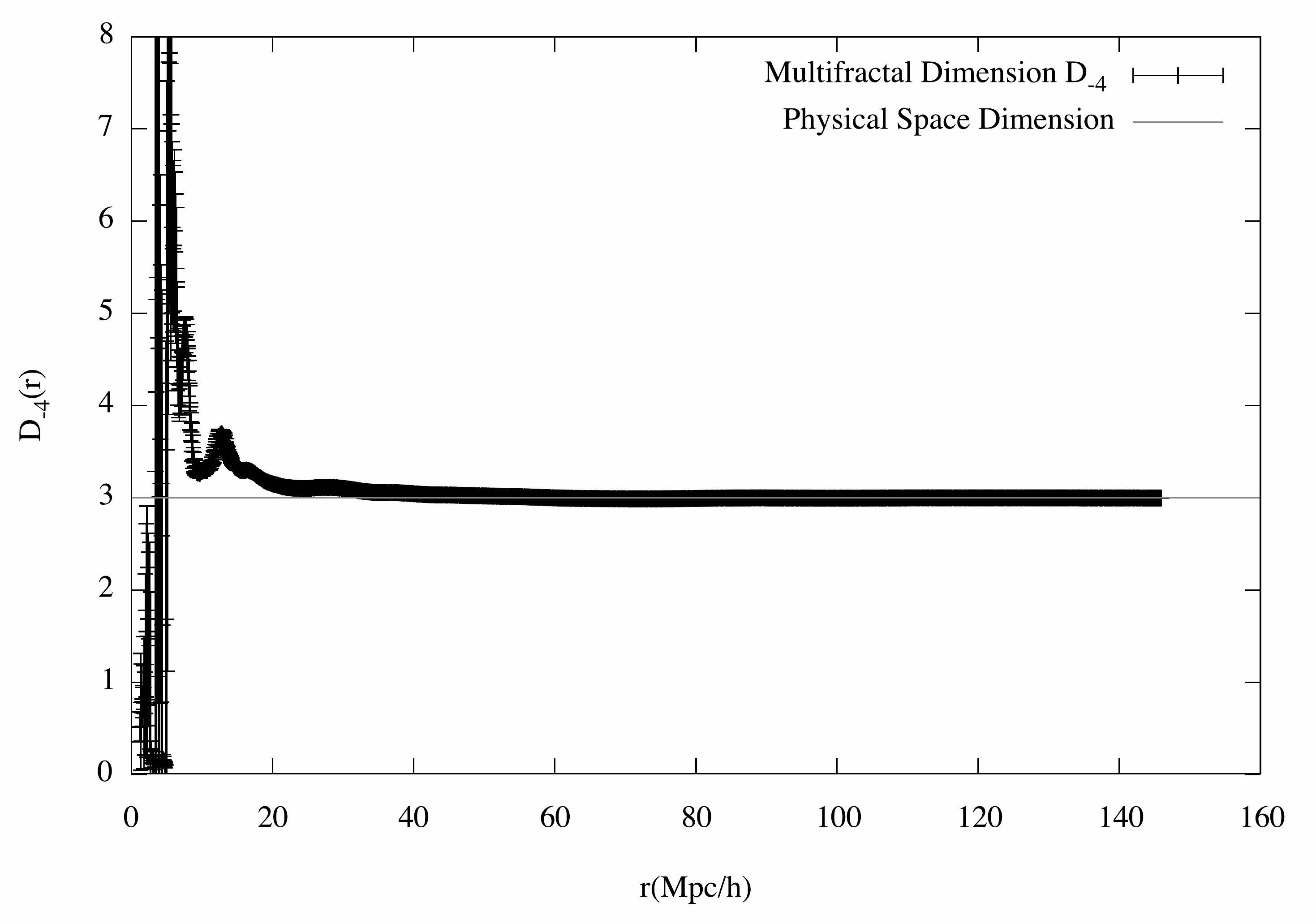}
         	\includegraphics[width=0.45\linewidth,clip]{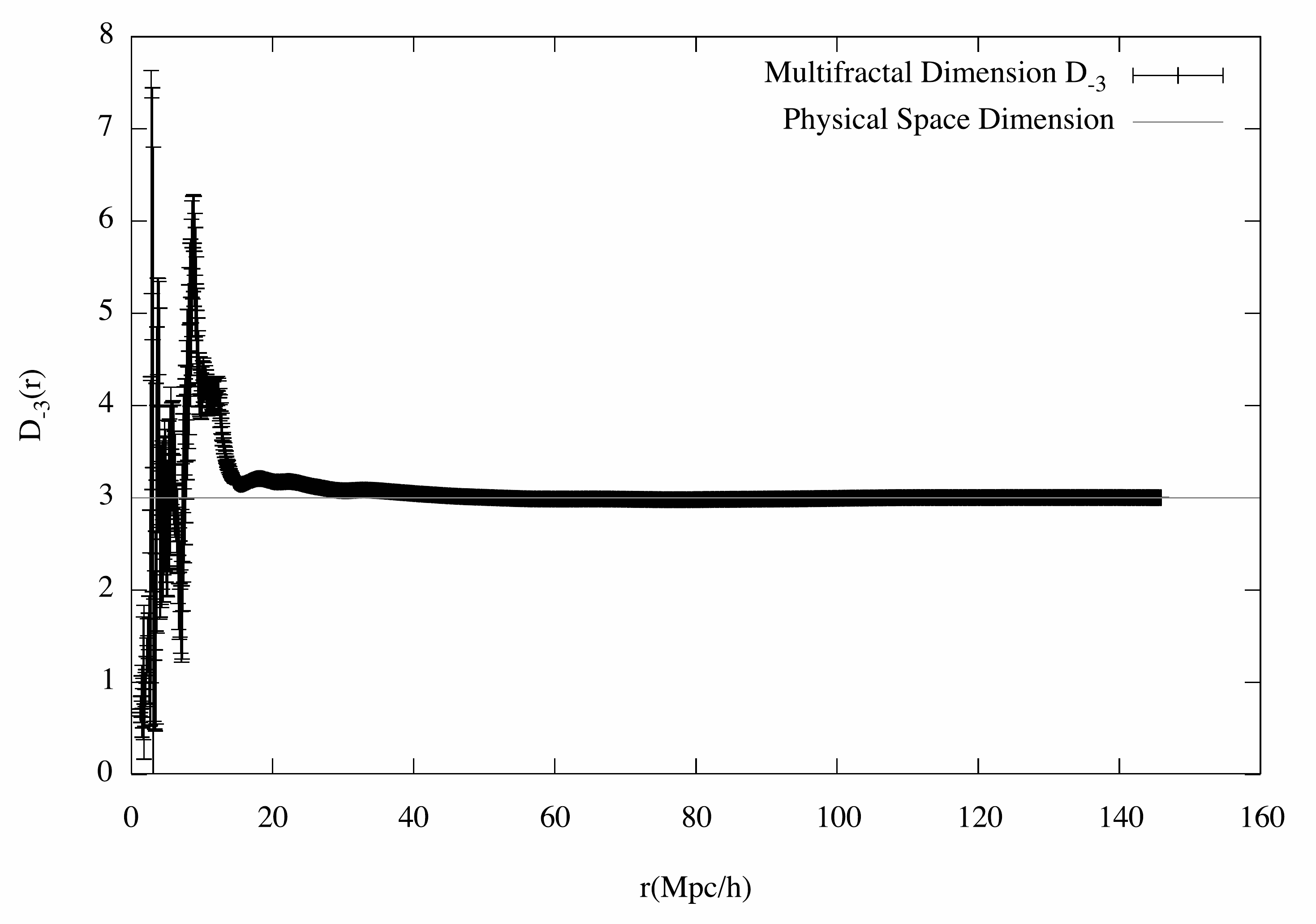}
	\includegraphics[width=0.45\linewidth,clip]{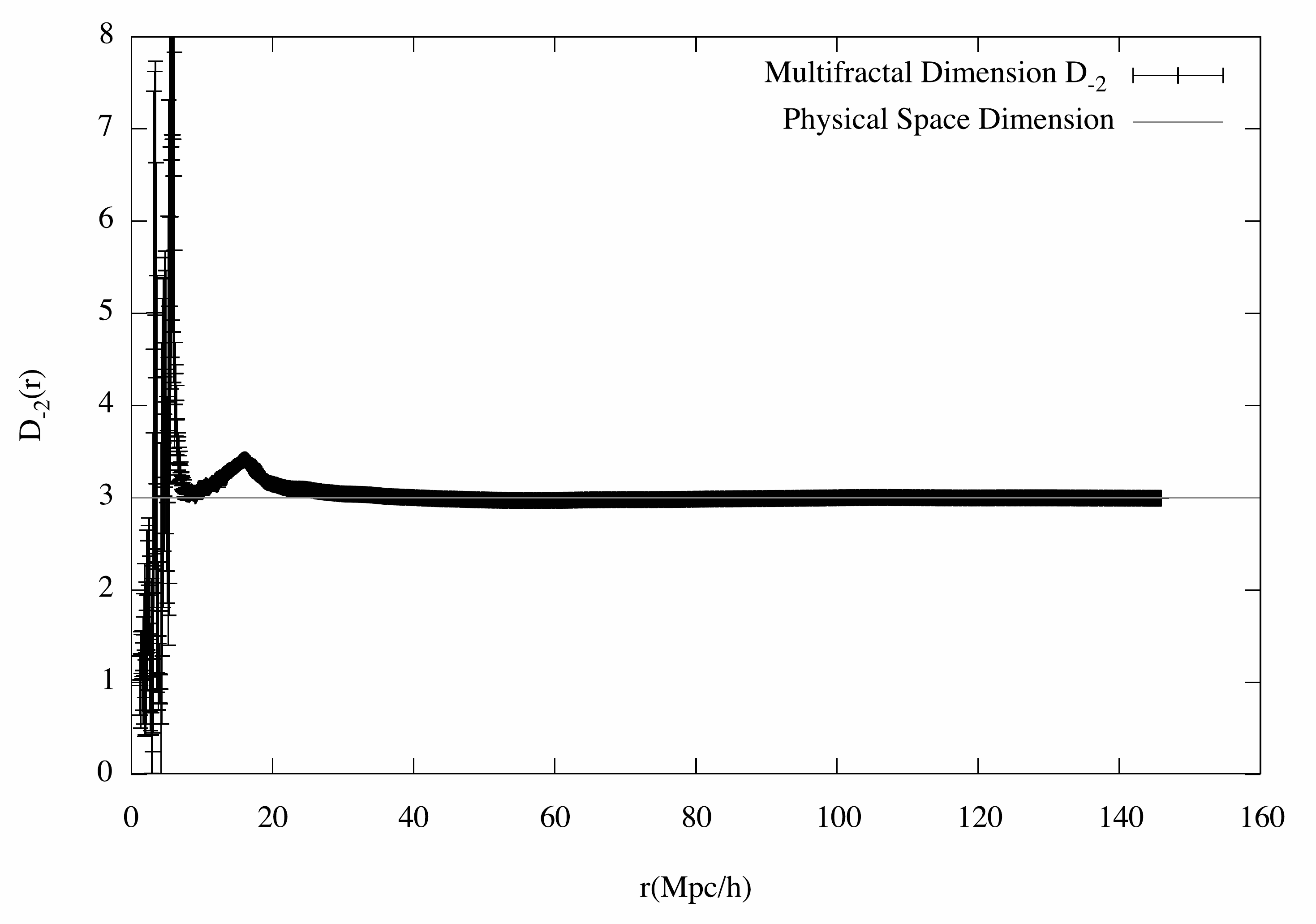}
 	\includegraphics[width=0.45\linewidth,clip]{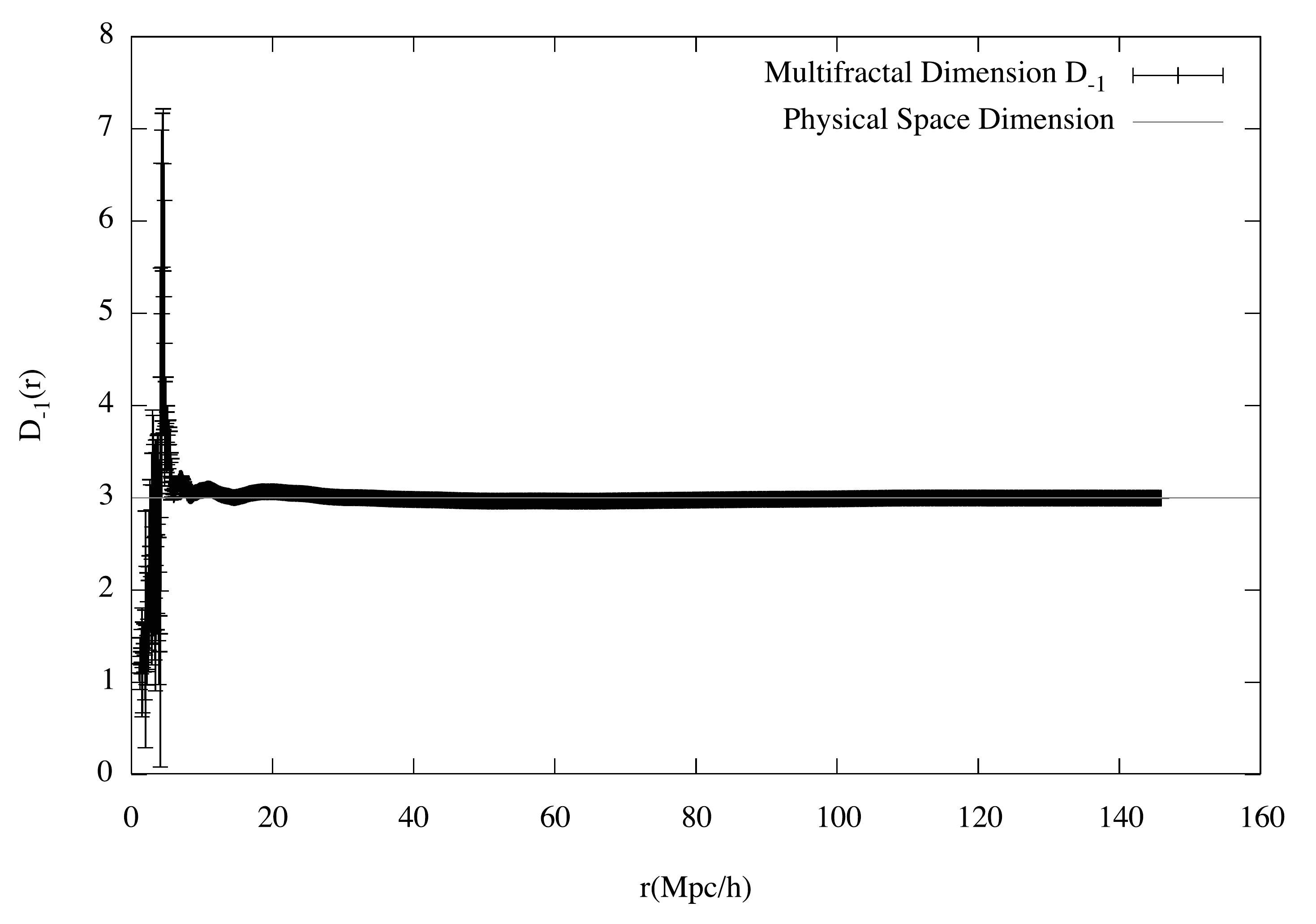}
         \includegraphics[width=0.45\linewidth,clip]{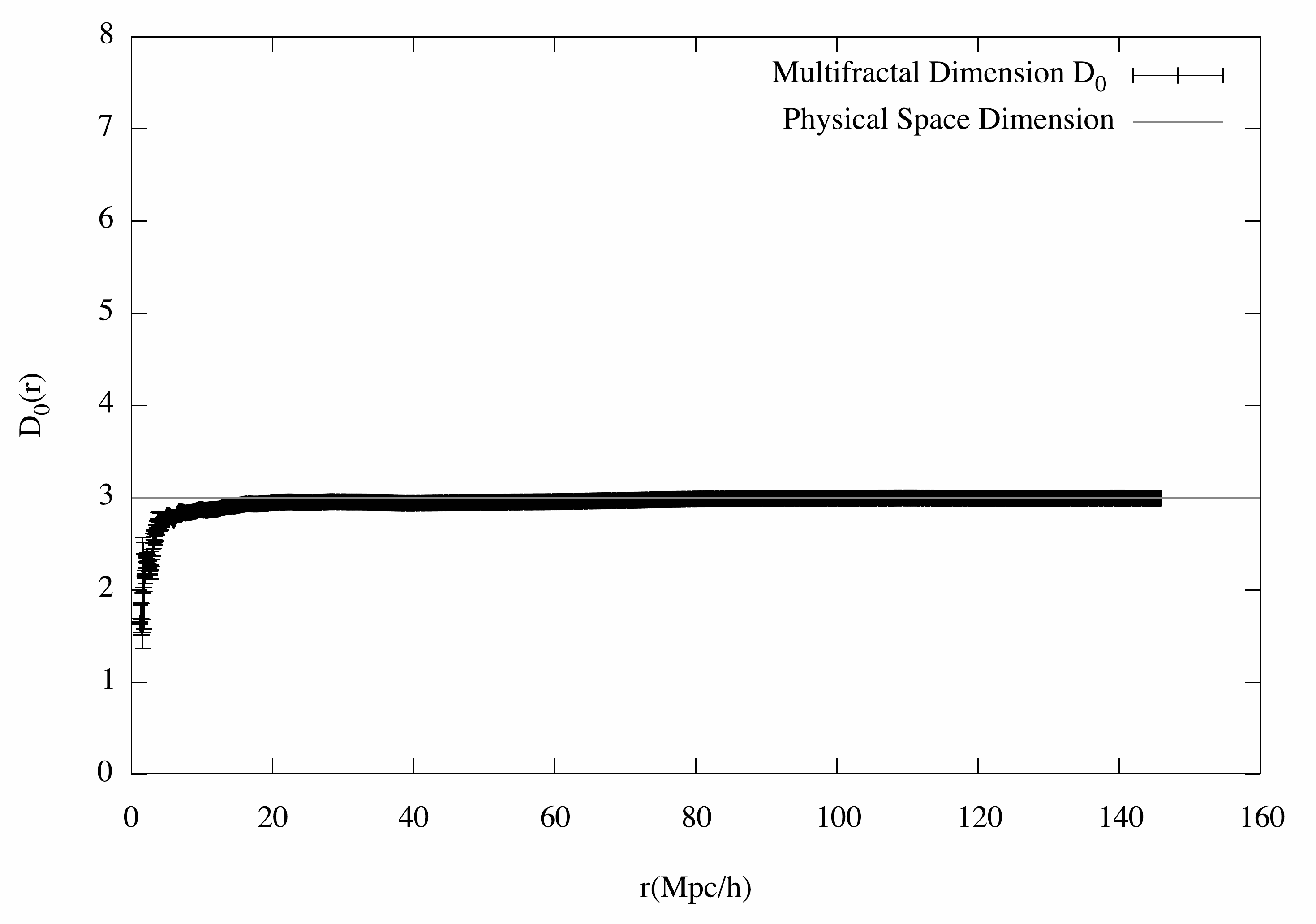}
	
	\caption{Multi-fractal Dimension Spectrum  $D_q(r)$ \textit{vs} radial distance $r$ for low density environments $-6 \leqslant q  \leqslant 0$ from the Millennium dark matter haloes. $\pm1\sigma$ error bars in the graphics, with 0.99 confidence level in t-student test.} 
	\label{Dqr1a}
\end{figure*}

From the generalised correlation integral, the sliding window technique is used to determine the fractal dimension spectrum. We find the multi-fractal dimension behaviour as a function of the radial distance for the structure parameter values $q < 1$ as shown in the Figure \ref{Dqr1a}. Besides, we calculate the corresponding $\chi^2$ test in each multi-fractal relation over all distance range where we made the calculations; the low $\chi^2$ values provide us confidence on the used method. The $\chi^2$ average and maximum values for each structure parameter are shown in Table \ref{table:chi2v}.

\begin{figure*}
	\includegraphics[width=0.45\linewidth,clip]{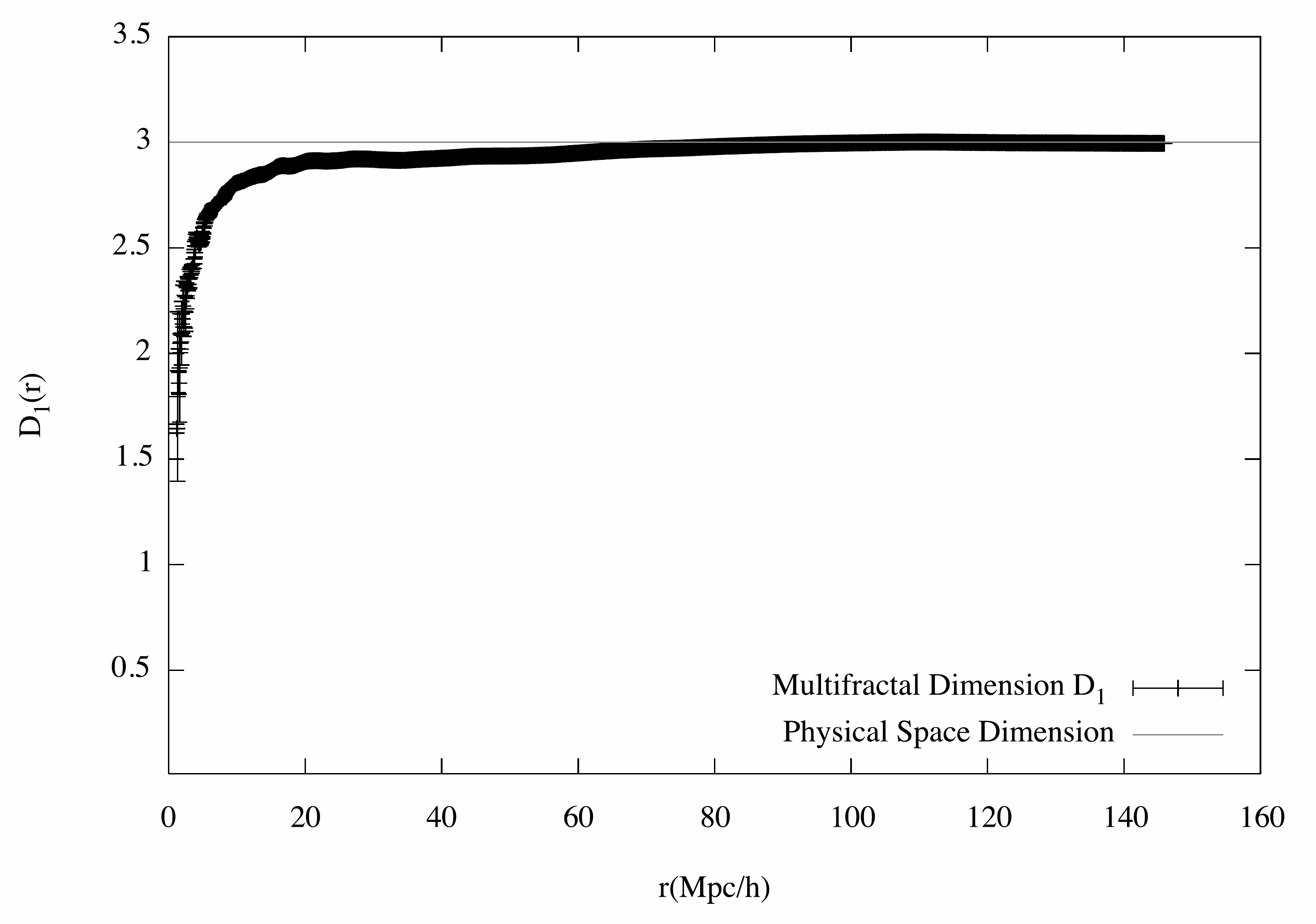}
	\includegraphics[width=0.45\linewidth,clip]{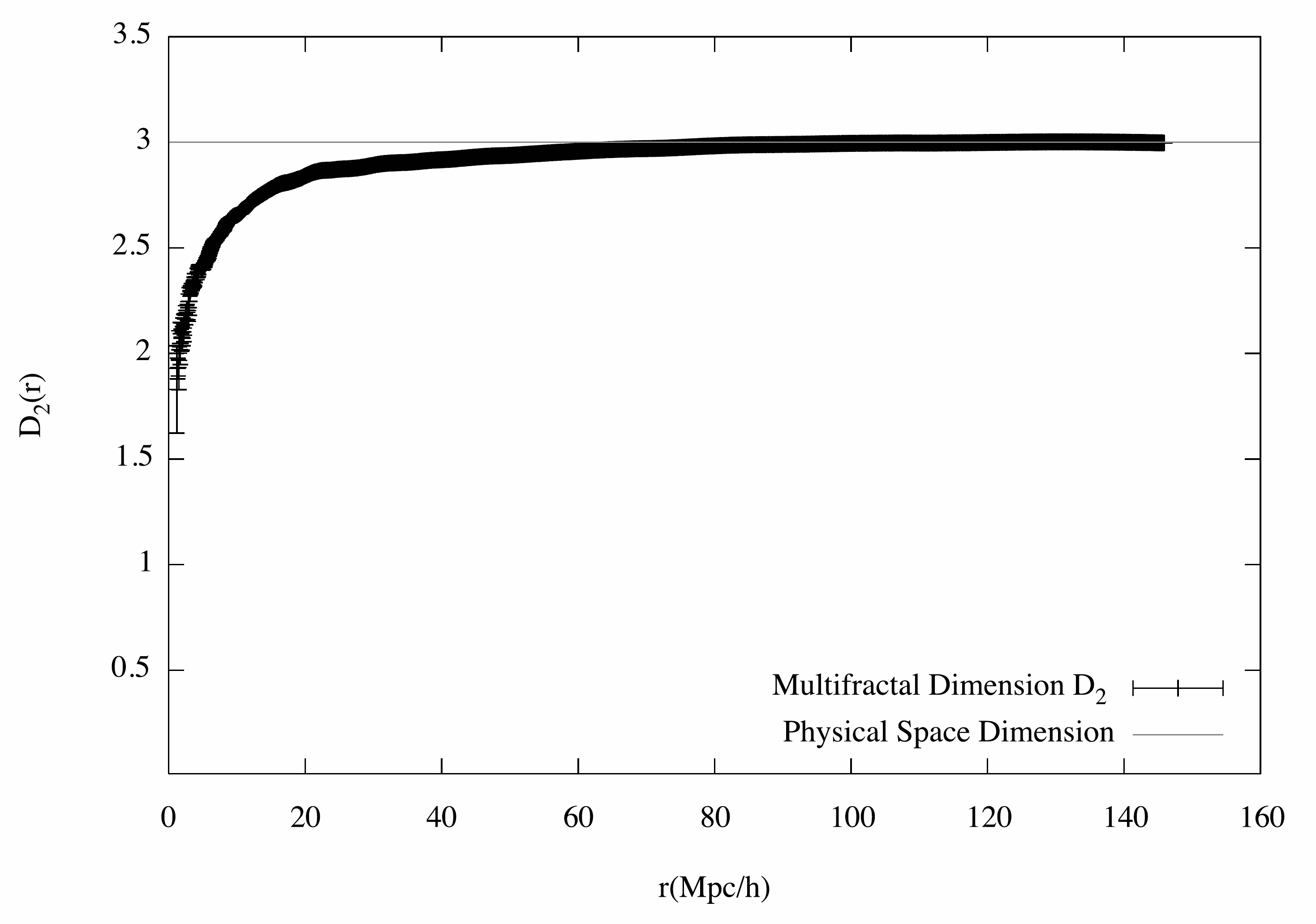}
	\includegraphics[width=0.45\linewidth,clip]{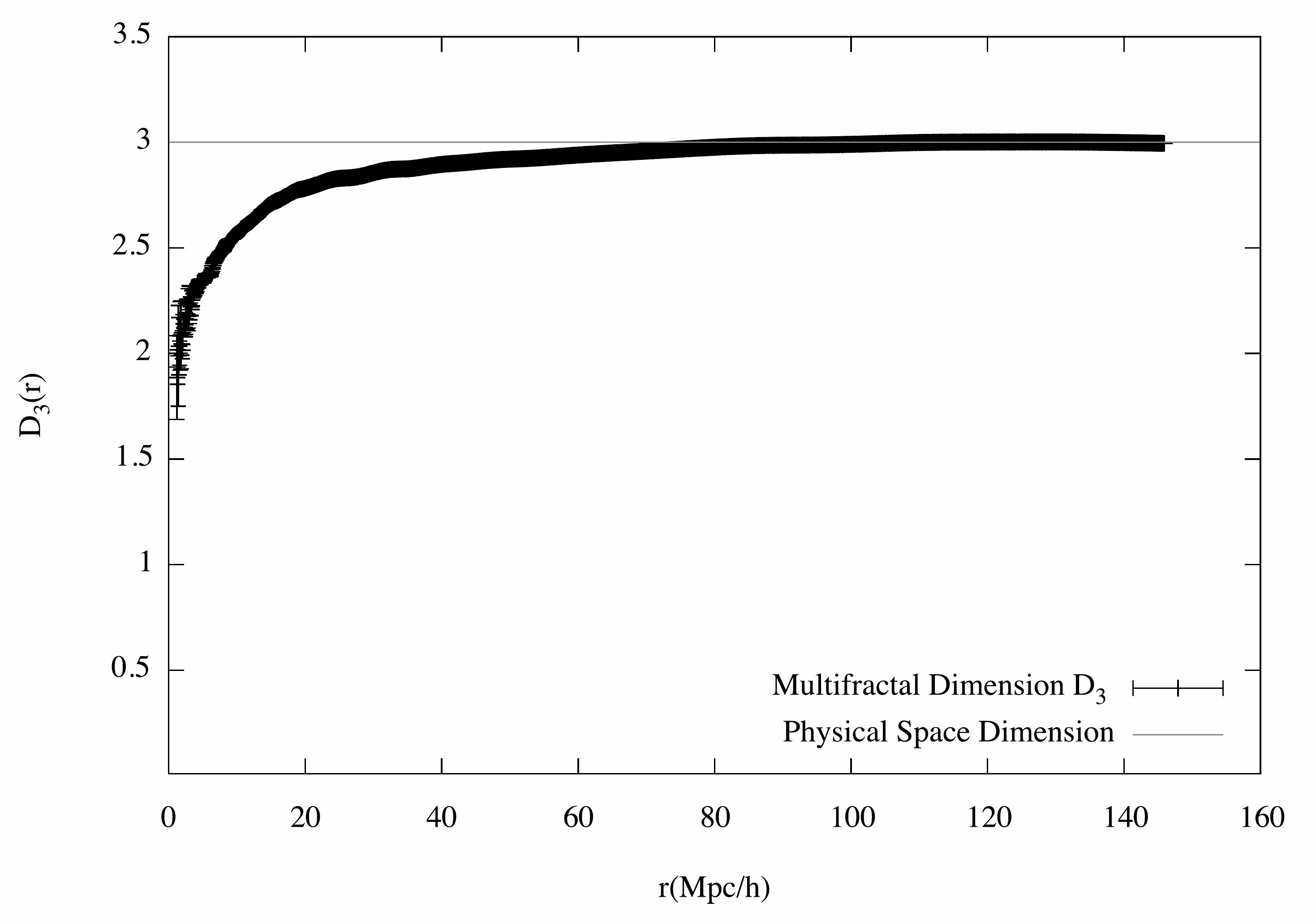}
	\includegraphics[width=0.45\linewidth,clip]{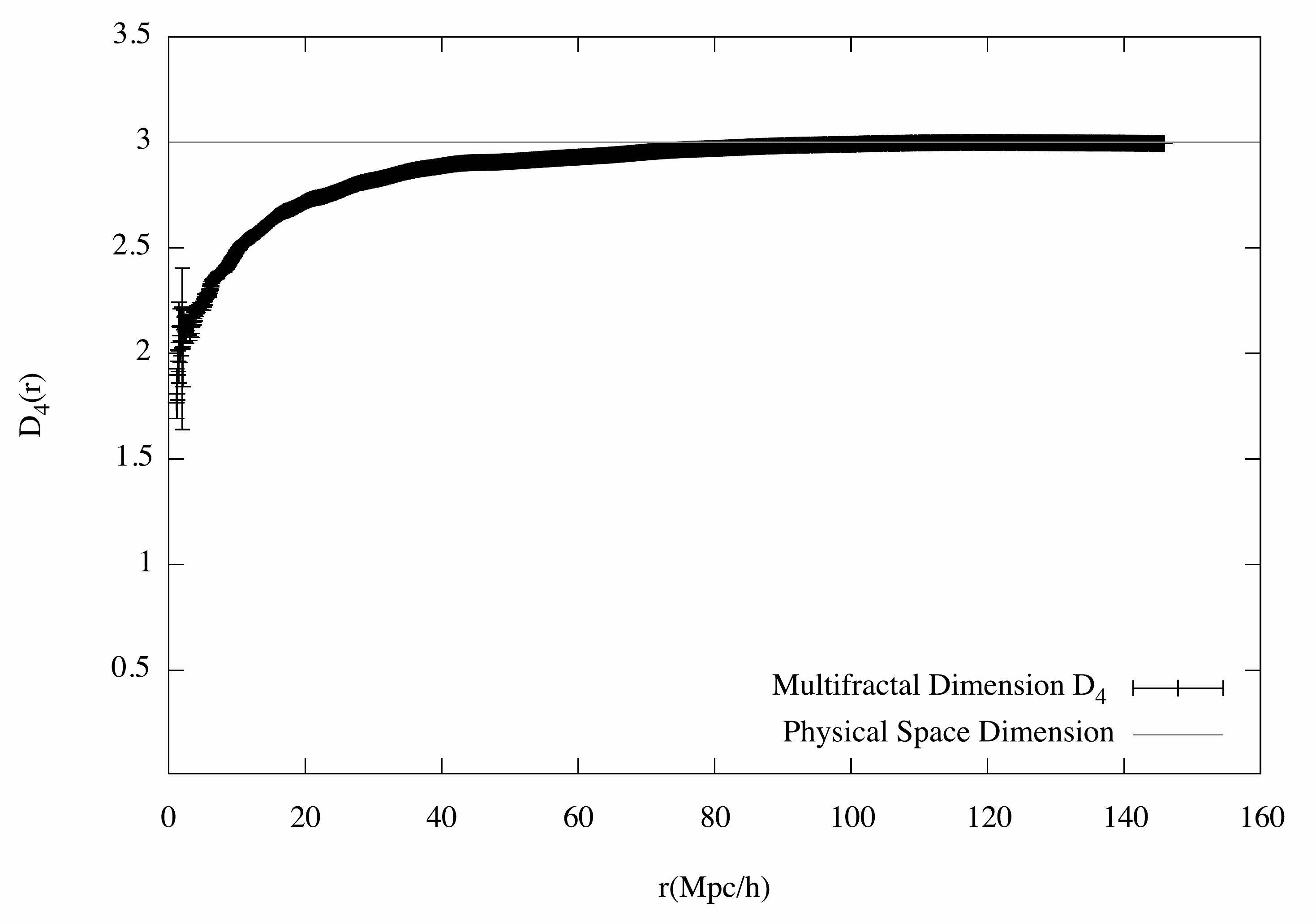}
	\includegraphics[width=0.45\linewidth,clip]{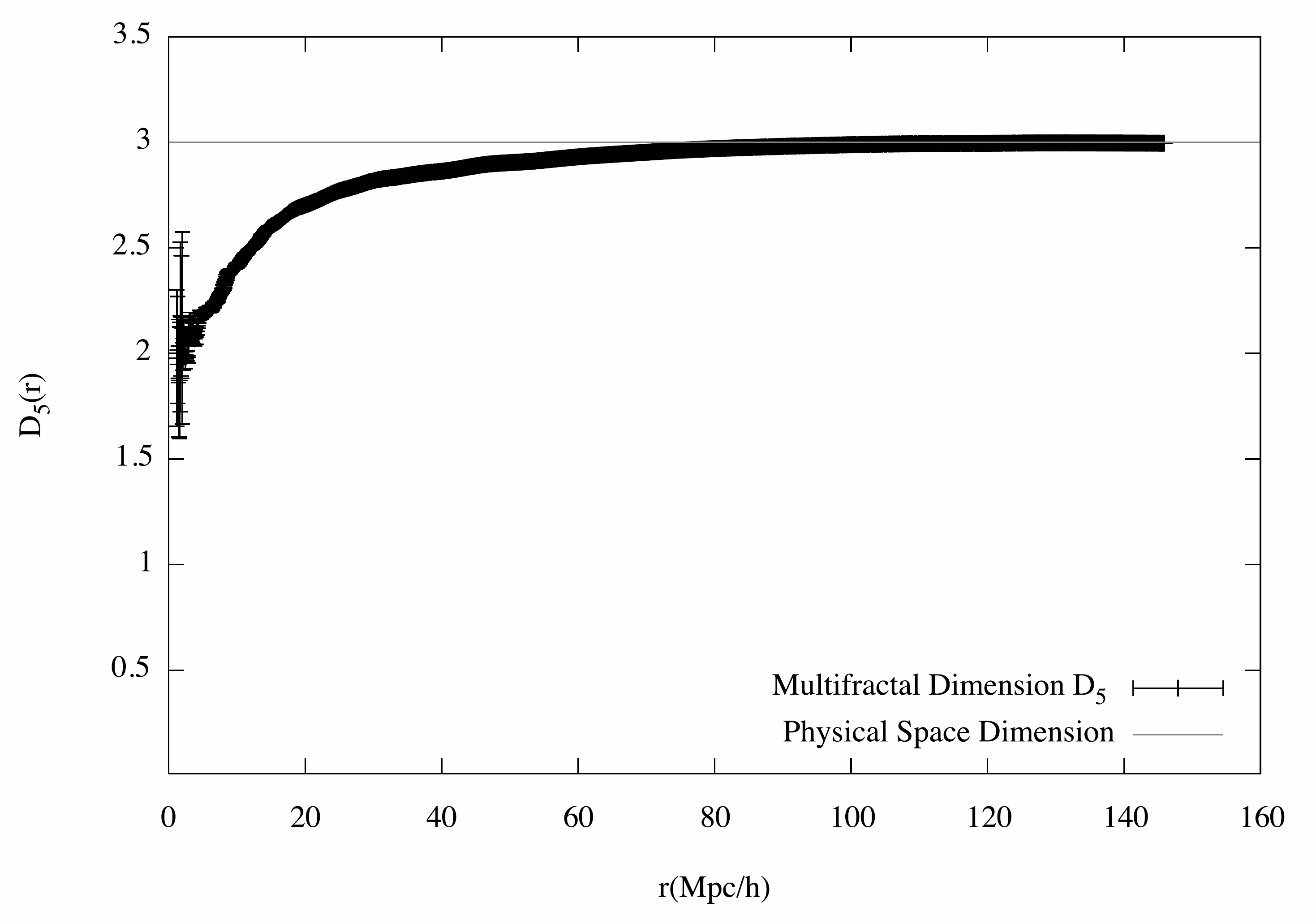}
	\includegraphics[width=0.45\linewidth,clip]{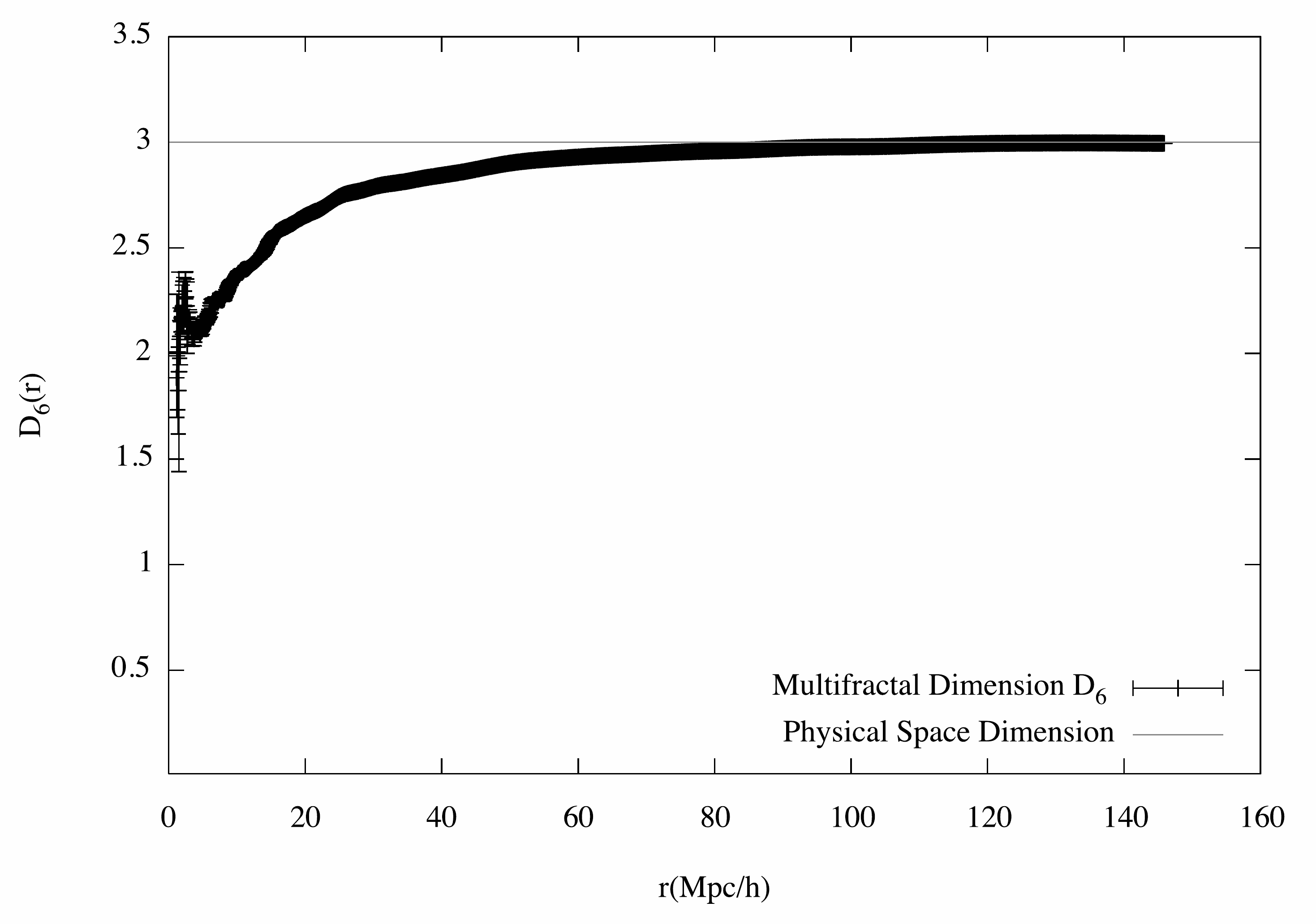}
	
	\caption{Multi-fractal dimension spectrum  $D_q(r)$ \textit{vs} radial distance $r$ for high density environments $ 1\leqslant q \leqslant 6$  from the Millennium dark matter haloes. $\pm1\sigma$ error bars in the graphics, with 0.99 confidence level in t-student test.} 
	
\label{Dqr2a}
\end{figure*}

For structure parameter values $q \geqslant 1$ the radial distance dependency is presented in Figure \ref{Dqr2a}. In this manner we covered the range of the structure parameter $-6 \leqslant q \leqslant 6$ in steps of one. The  $\chi^2$ test was calculated in the same manner as for the multi-fractal relations with $q < 1$. There are low $\chi^2$ values over all radial distances, See Table \ref{table:chi2v}.

\begin{table*}
         	\centering 
	\begin{tabular}{ccccccccccccccccc}
 	\hline
	$q$ & $-6$ & $-5$ & $-4$ & $-3$ & $-2$ & $-1$ & $0$\\
	\hline
	$<\chi^2>$ & $1.9\times10^{-3}$ & $8.8\times10^{-4}$ & $2.5\times10^{-3}$ & $1.2\times10^{-3}$ & $1.5\times10^{-3}$ & $3.1\times10^{-4}$ & $2.5\times10^{-6}$\\
	\hline
	$\chi^2 _{max}$ & $4.5\times10^{-1}$ & $1.0\times10^{-1}$ & $2.9\times10^{-1}$ & $6.7\times10^{-2}$ & $2.1\times10^{-1}$ & $7.1\times10^{-2}$ & $1.3\times10^{-4}$\\
	\hline
	$q$ & $1$ & $2$ & $3$ & $4$ & $5$ & $6$ \\
	\hline
	$<\chi^2>$ & $1.6\times10^{-6}$ & $1.6\times10^{-6}$ & $2.6\times10^{-6}$ & $3.1\times10^{-6}$ & $3.5\times10^{-6}$ & $4.7\times10^{-6}$\\
	\hline
	$\chi^2 _{max}$ & $9.0\times10^{-5}$ & $2.8\times10^{-5}$ & $2.8\times10^{-5}$ & $4.7\times10^{-5}$ & $6.9\times10^{-5}$ & $9.4\times10^{-5}$\\
	\hline
	\end{tabular}
	
\caption{$\chi^2$ mean and maximum values obtained for each multi-fractal dimension $D_q$. Top: low density environments $-6 \leqslant q  \leqslant 0$, bottom:  high density environments $ 1\leqslant q \leqslant 6$. For all the $D_q$ analysed  the $\chi^2$ values show a peak at scales smaller than $\approx 15$ Mpc/h and very low, almost constant, values at larger scales.}
\label{table:chi2v}
\end{table*}

In order to detect the transition to homogeneity in terms of the fractal dimension, we use the relationship:   
\begin{equation}
 \mid D_q-3 \mid / 3 < \epsilon, 
\end{equation}
where $\epsilon$ is a fiducial value which is calculated from the average of the relative dispersion $\sigma/D_q$.  In Table \ref{table:Rh} we present the shortest radial distances for which this condition for the transition to homogeneity is fulfilled.

\begin{table*}
         	\centering 
	\begin{tabular}{ccccccccccccccccc}
	\hline
	$q$ & $1$ & $2$ & $3$ & $4$ & $5$ & $6$ \\
	\hline
	$\epsilon$ & $1.8 \times10^{-3} $ & $8.4 \times10^{-4}$ & $1.2 \times10^{-3}$ & $1.3 \times10^{-3} $ & $1.8 \times10^{-3}$ & $ 1.7 \times10^{-3}$ \\
	\hline
	$R_{h}[Mpc/h]$ & $101.1$ & $108.9$ & $107.2$ & $108.7$ & $117.7$ & $122.5$ \\
	\hline
	\end{tabular}
	
\caption{Scale of homogeneity transition $R_{h}$  for every fractal dimension in overdenses environments $1\leqslant q \leqslant 6$. The $\epsilon$ value is calculated from the relative dispersion of the fractal dimension  }
\label{table:Rh}
\end{table*} 

\begin{figure*}
	\includegraphics[width=0.45\linewidth,clip]{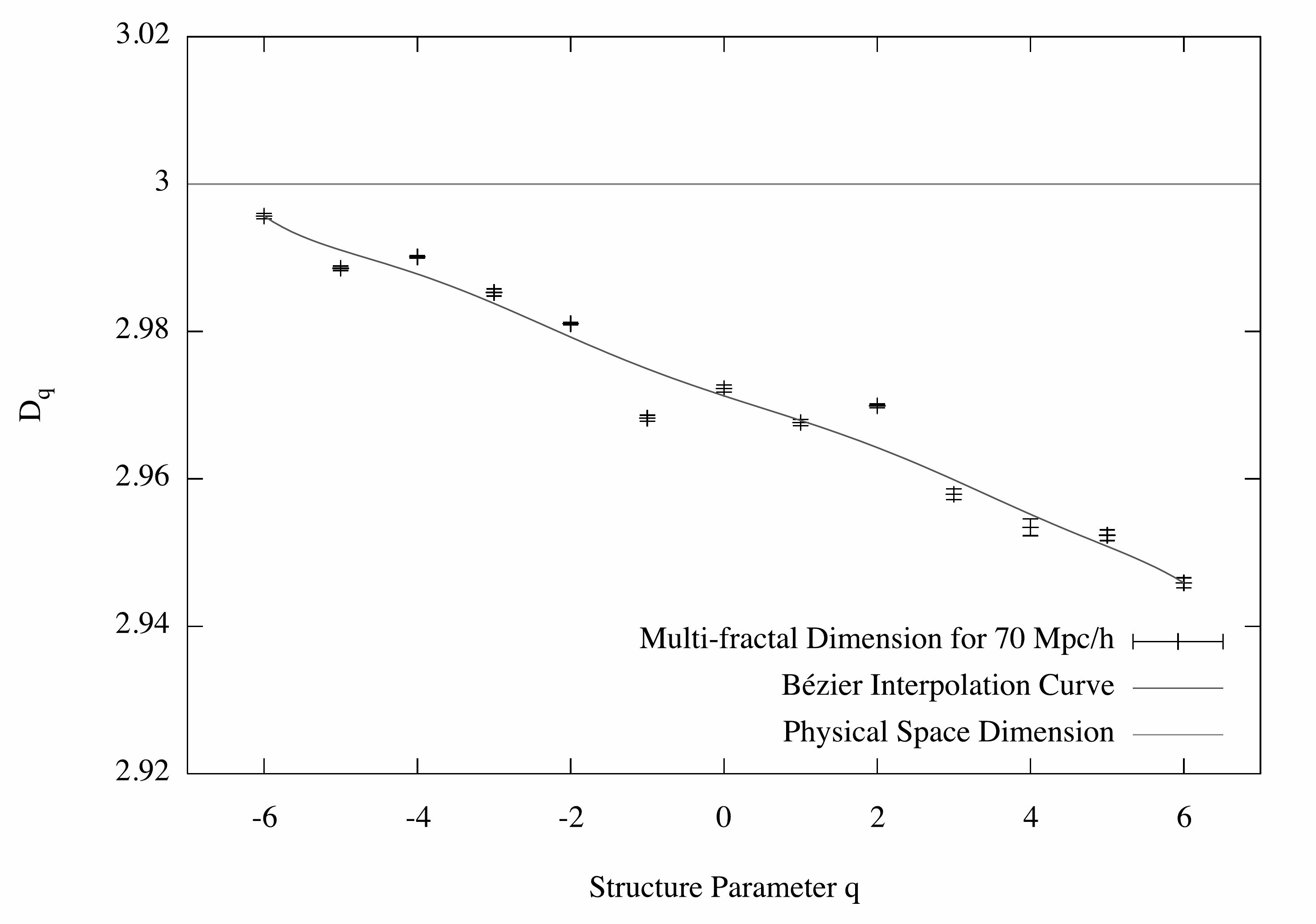}
	\includegraphics[width=0.45\linewidth,clip]{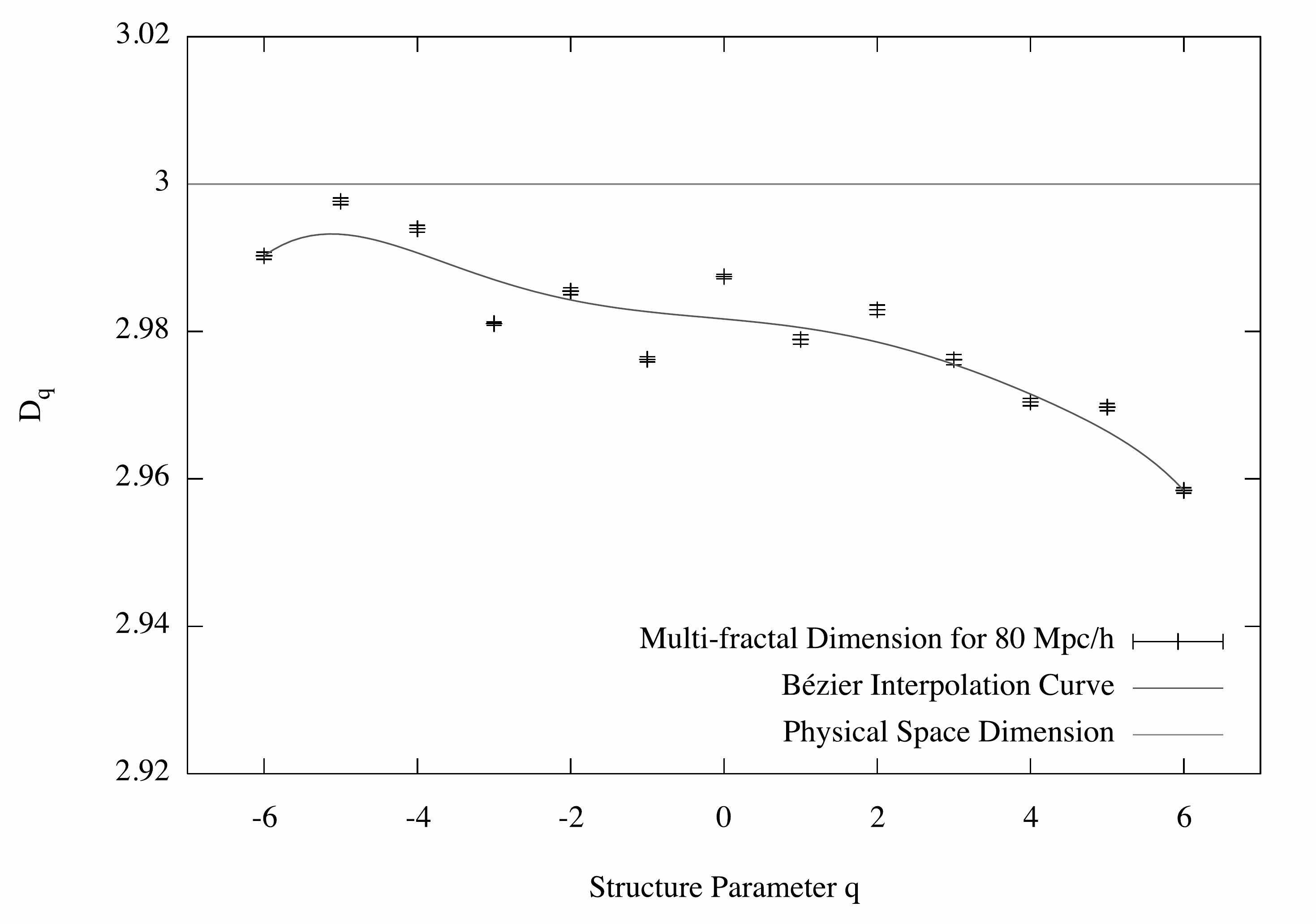}
	\includegraphics[width=0.45\linewidth,clip]{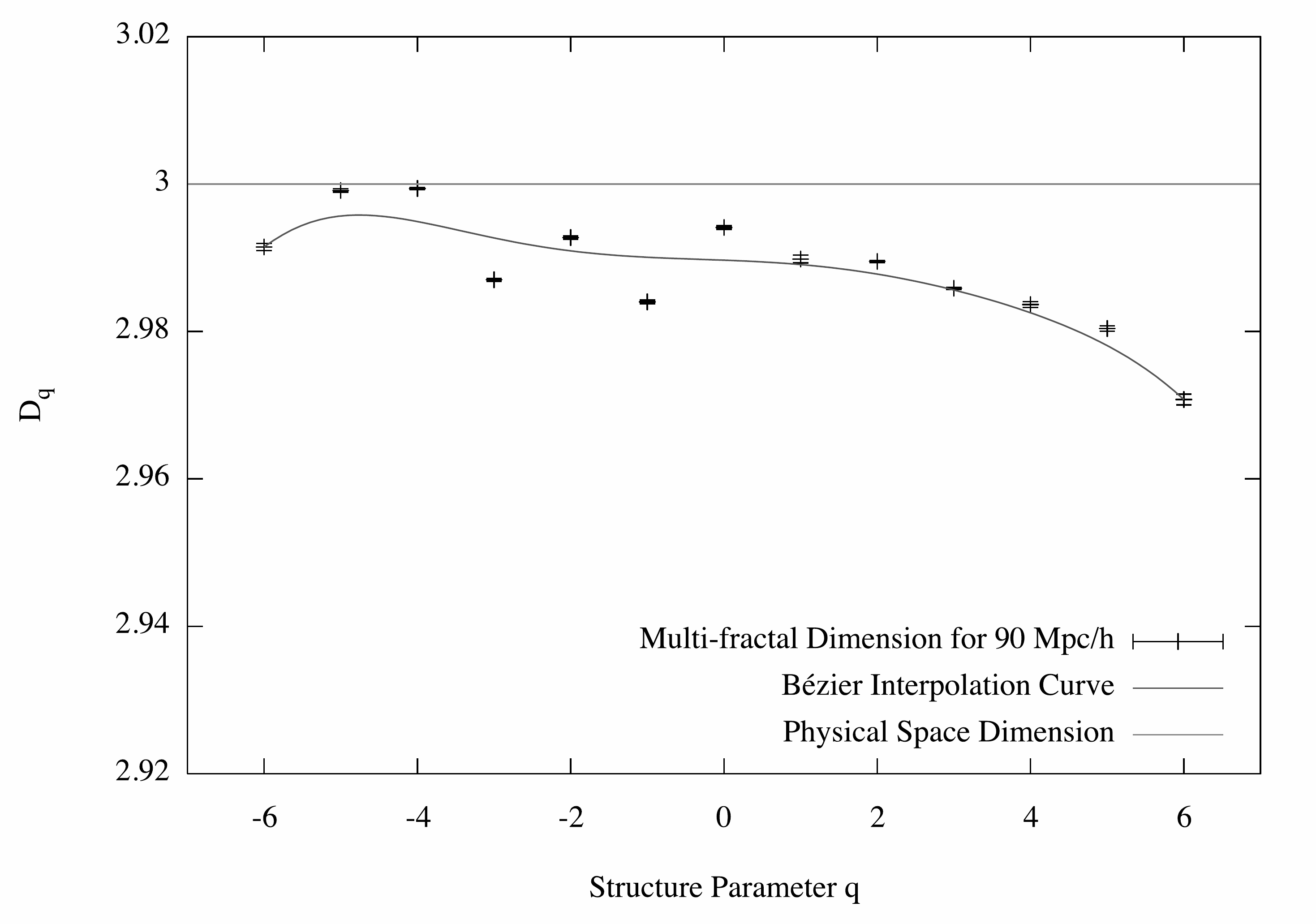}
         \includegraphics[width=0.45\linewidth,clip]{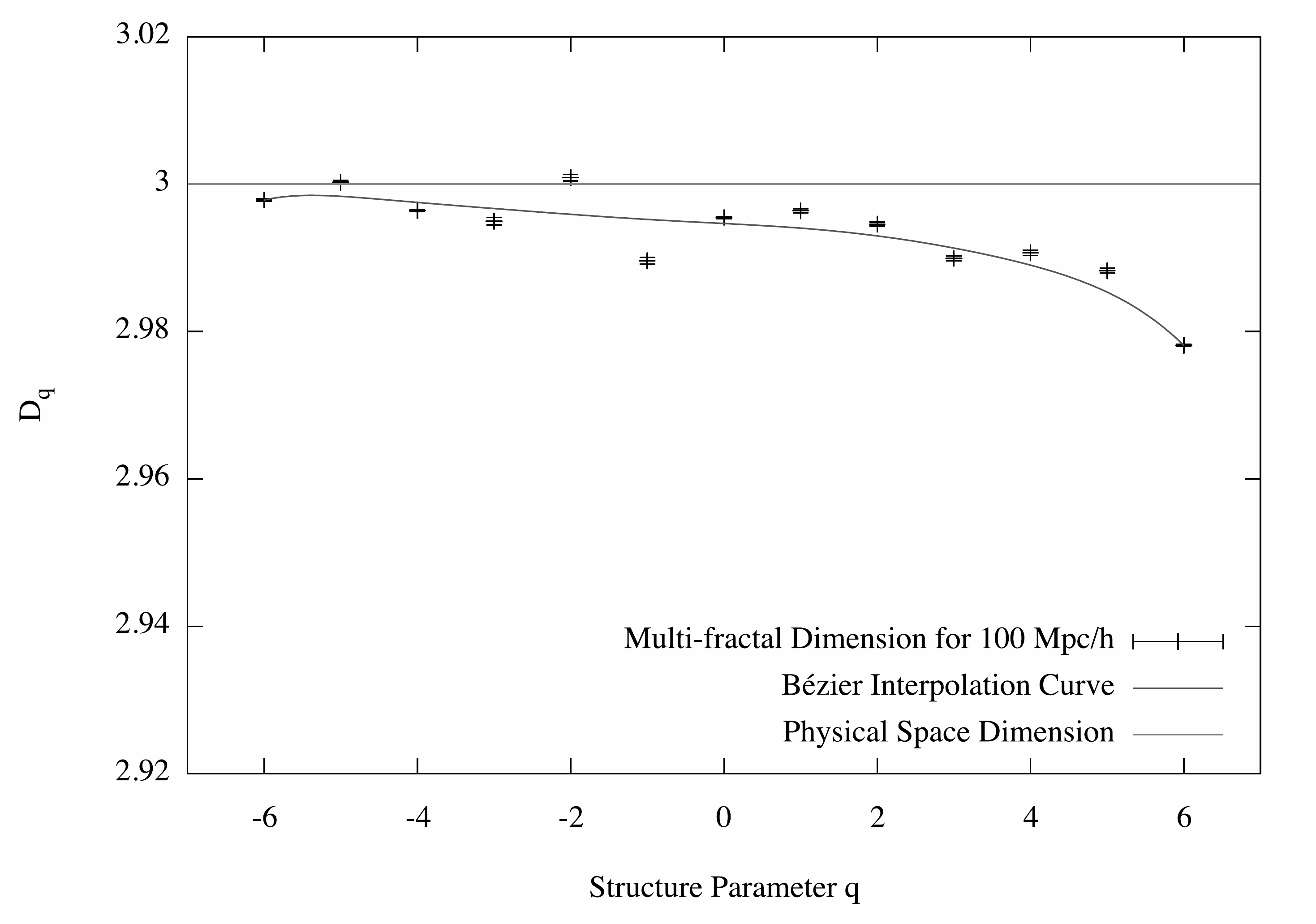}
          \includegraphics[width=0.45\linewidth,clip]{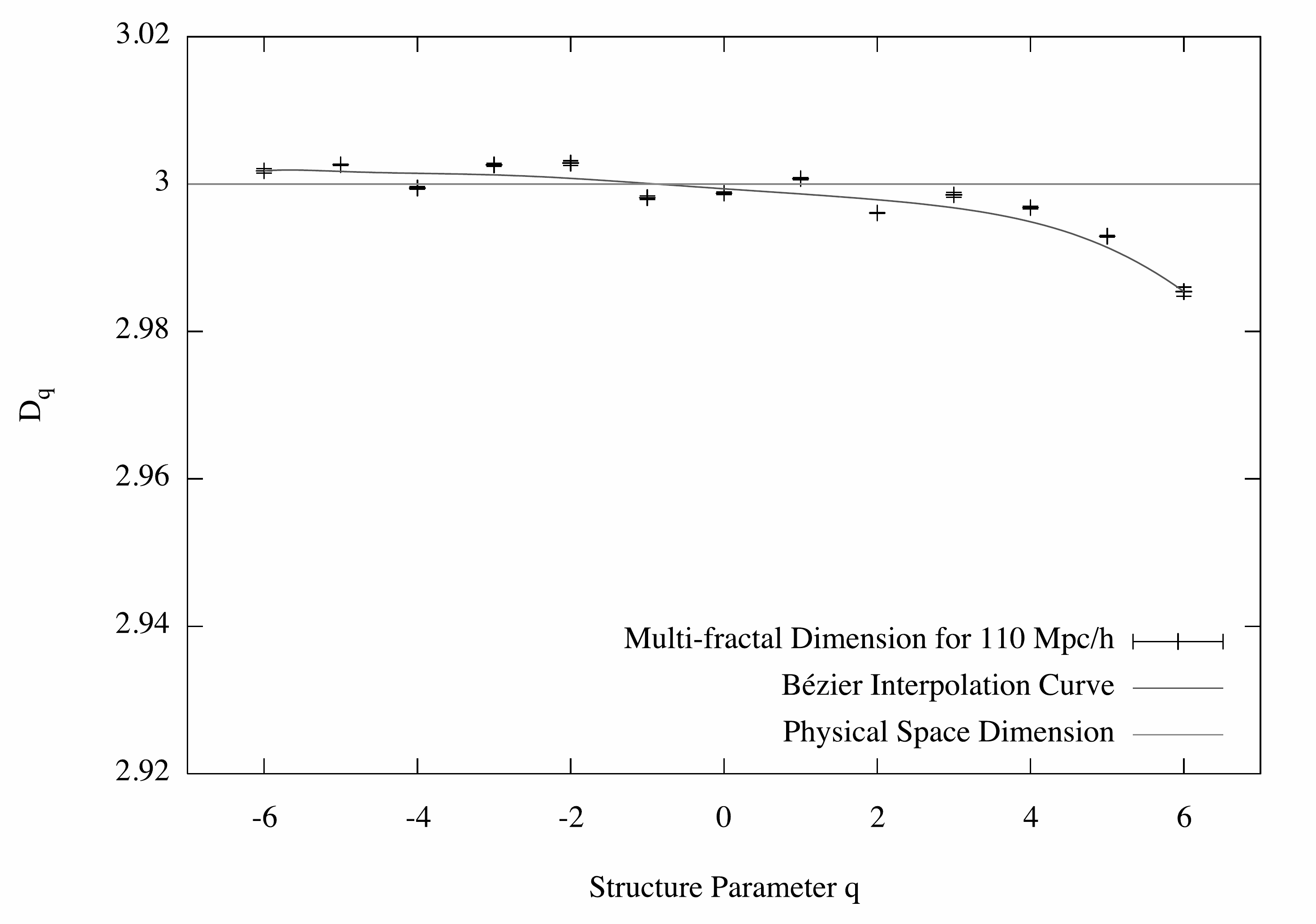}
         \includegraphics[width=0.45\linewidth,clip]{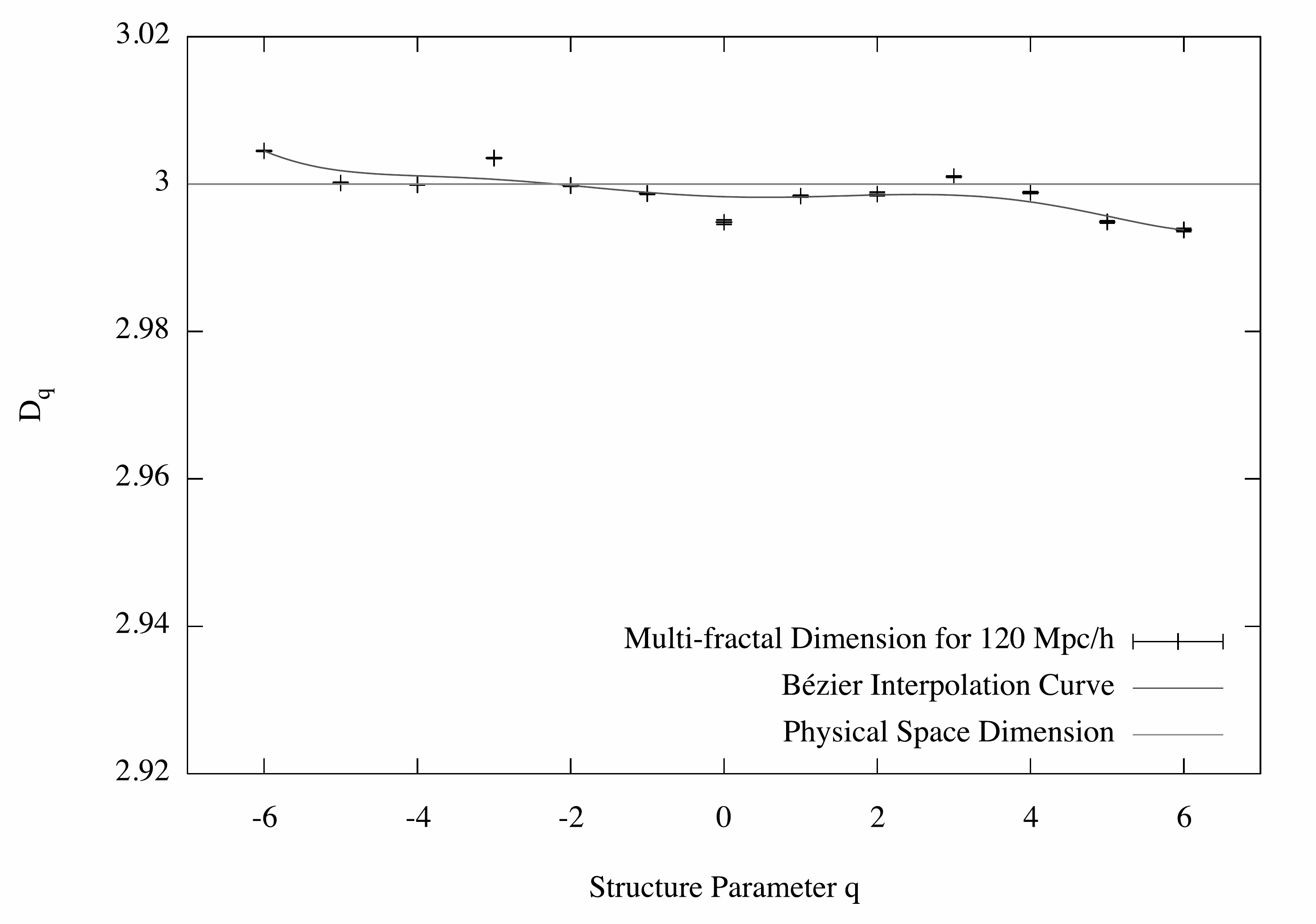} 
         \includegraphics[width=0.45\linewidth,clip]{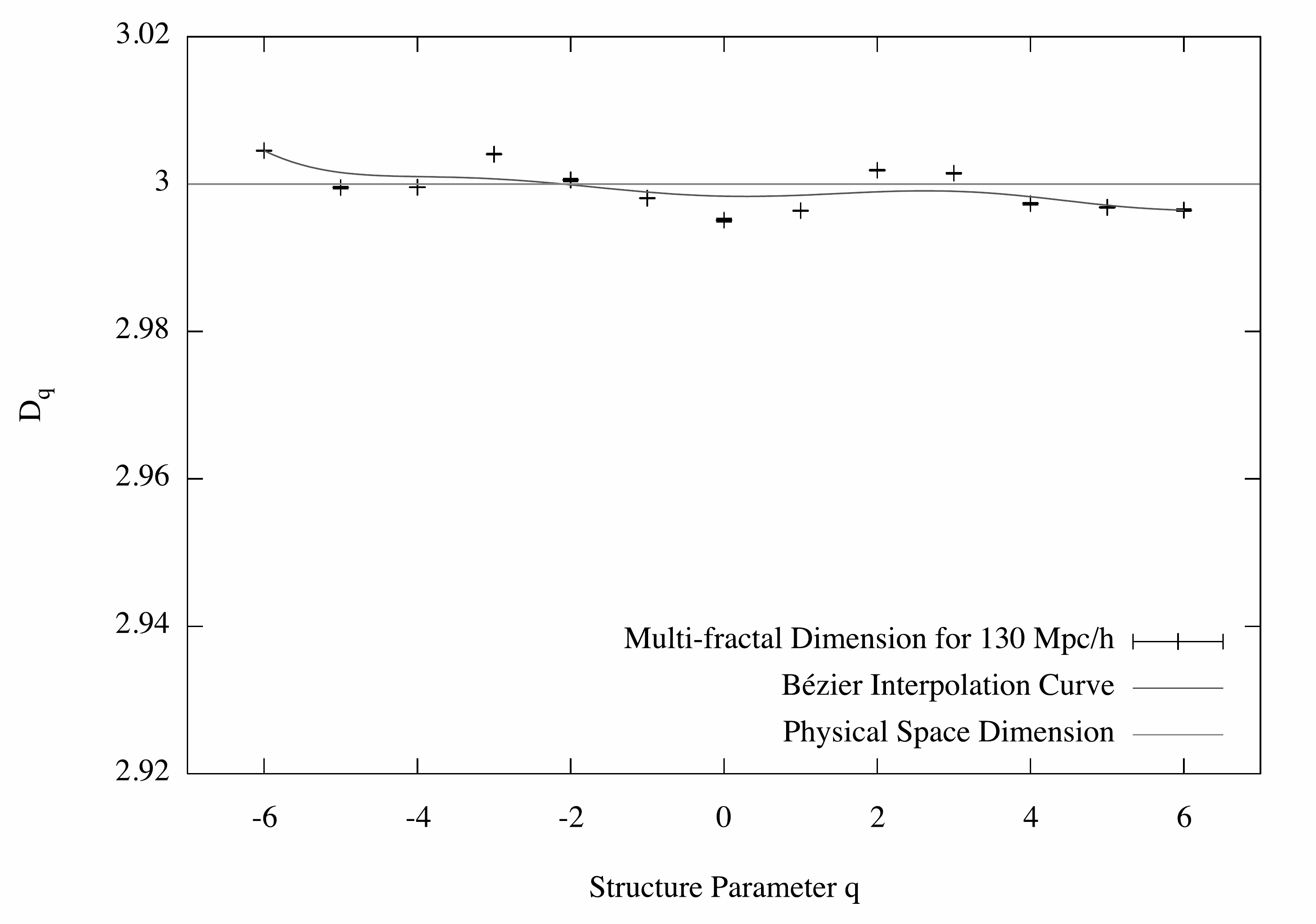} 
         \includegraphics[width=0.45\linewidth,clip]{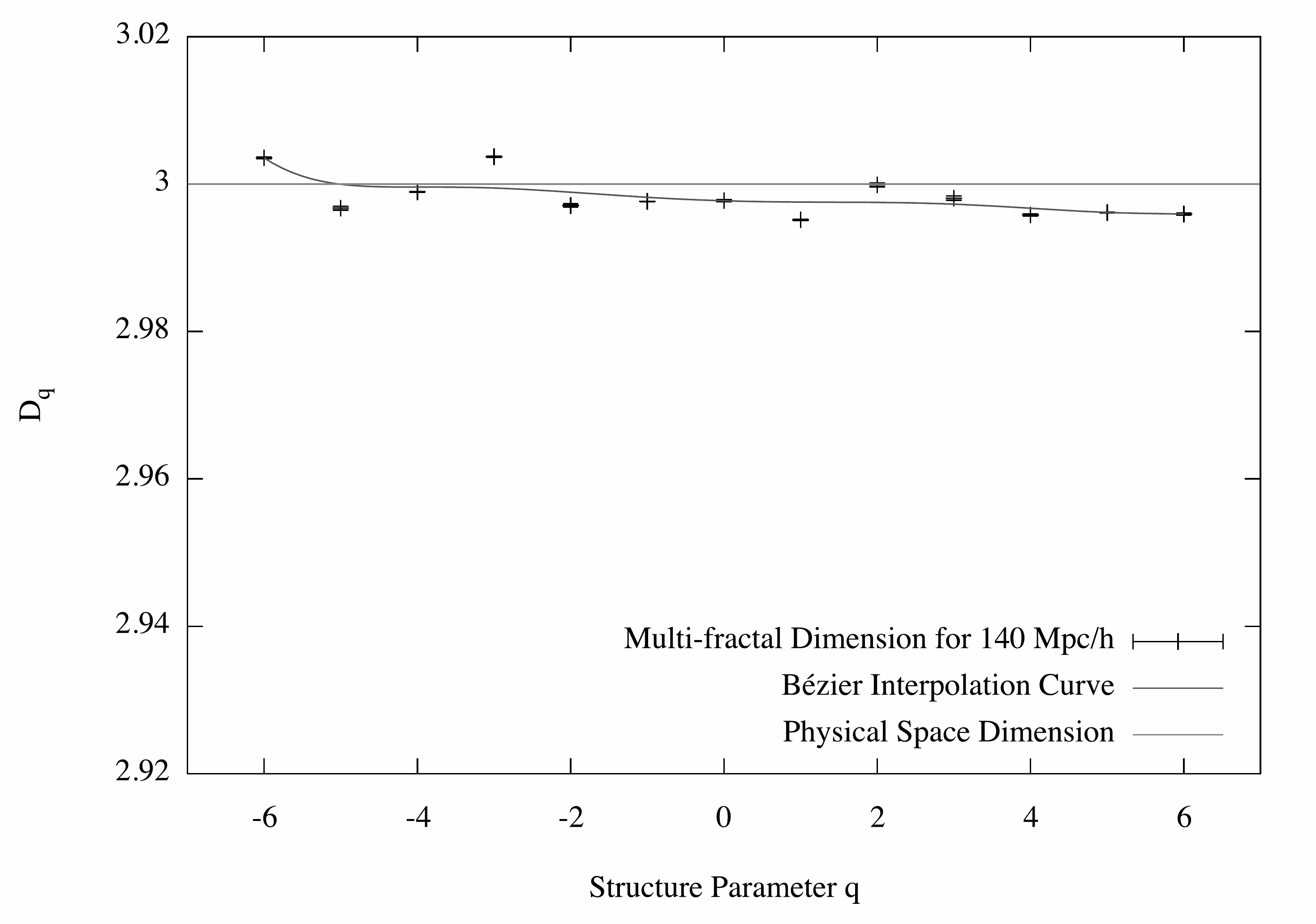} 
         \caption{Spectrum of the Multi-fractal dimension $D_q(r)$ as a function of structure parameter $q$ in the range $[-6,6]$ for Millennium simulation dark matter haloes. Homogeneity transition around 120 Mpc/h is determined. The solid line is an interpolation by the B\'ezier method.}
	\label{Dqrq}
\end{figure*}

This results are confirmed by the relationship between the multi-fractal dimension and the spectral parameter $q$, as shown in Figure \ref{Dqrq}.

Another possible definition for transition to homogeneity is: 

\begin{equation}
{2{[C_2(r_1)/V(r_1)-C_2(r_2)/V(r_2)]} \over {[C_2(r_1)/V(r_1)+C_2(r_2)/V(r_2)]} }< \epsilon, 
\end{equation}
$\forall  r_1,r_2 > R_h$, where $R_h$ is the homogeneity scale, and $\epsilon$ is a fiducial value. If  the same value for epsilon as above is used, then the transition to homogeneity appears at scales near 20 Mpc/h, which do not agree with the scales of transition to homogeneity we found from the multi-fractal spectrum or from the minimal values of the lacunarity spectrum.  Therefore, either this definition for transition to homogeneity is less appropriate than the former one or the value of epsilon should be much more smaller than $10^{-3}$ . In the latter case, it is not clear what fiducial value should be assigned to $\epsilon$ and how to estimate it.

To corroborate this result and because the fractal dimension does not indicate in what form the dark matter haloes set is filling the whole space, the analysis is completed with the calculation of the spectrum of lacunarity for overdenses regions, as shown below in Figure \ref{Phiqr}.

\begin{figure*}
	\includegraphics[width=0.45\linewidth,clip]{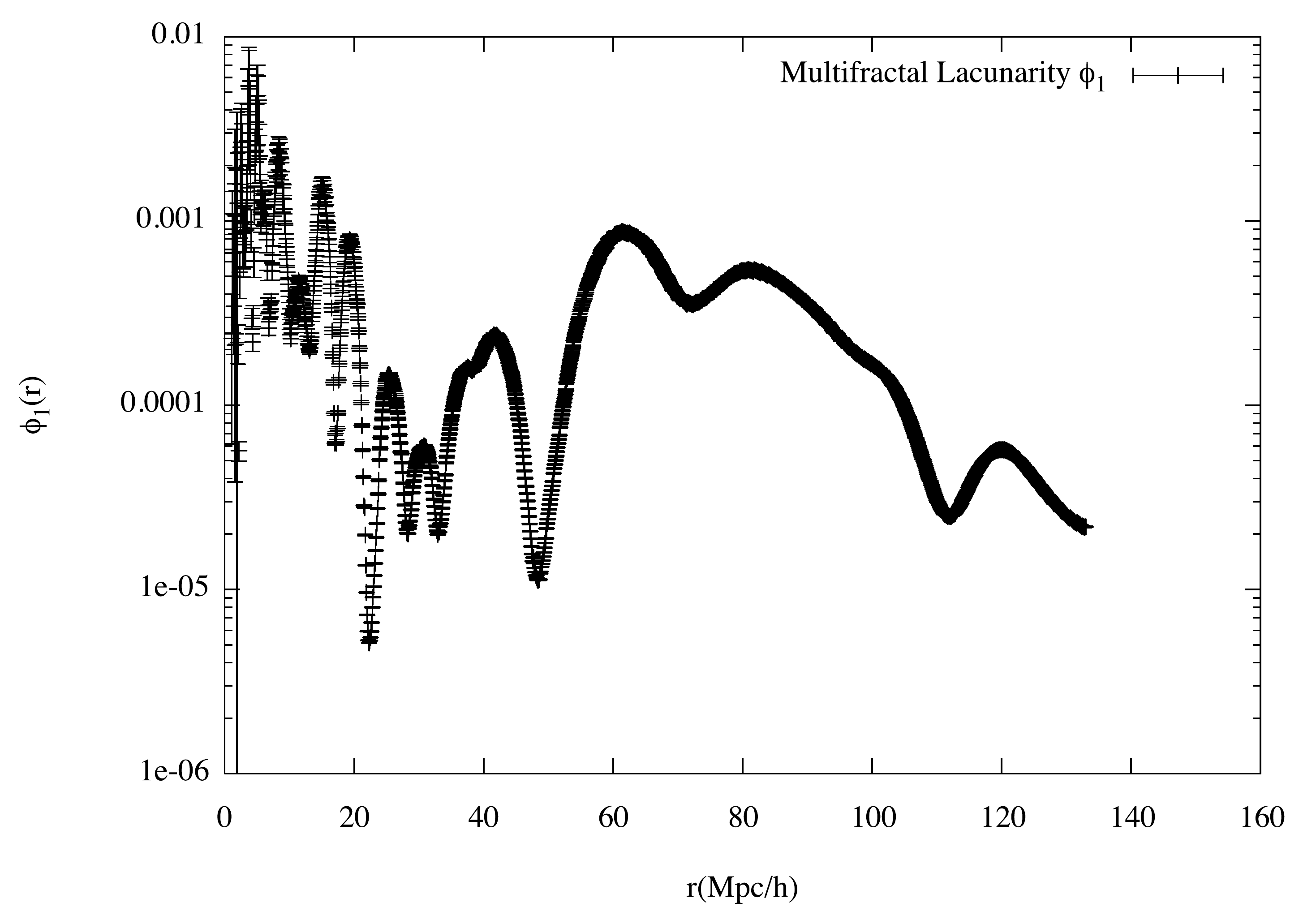}
	\includegraphics[width=0.45\linewidth,clip]{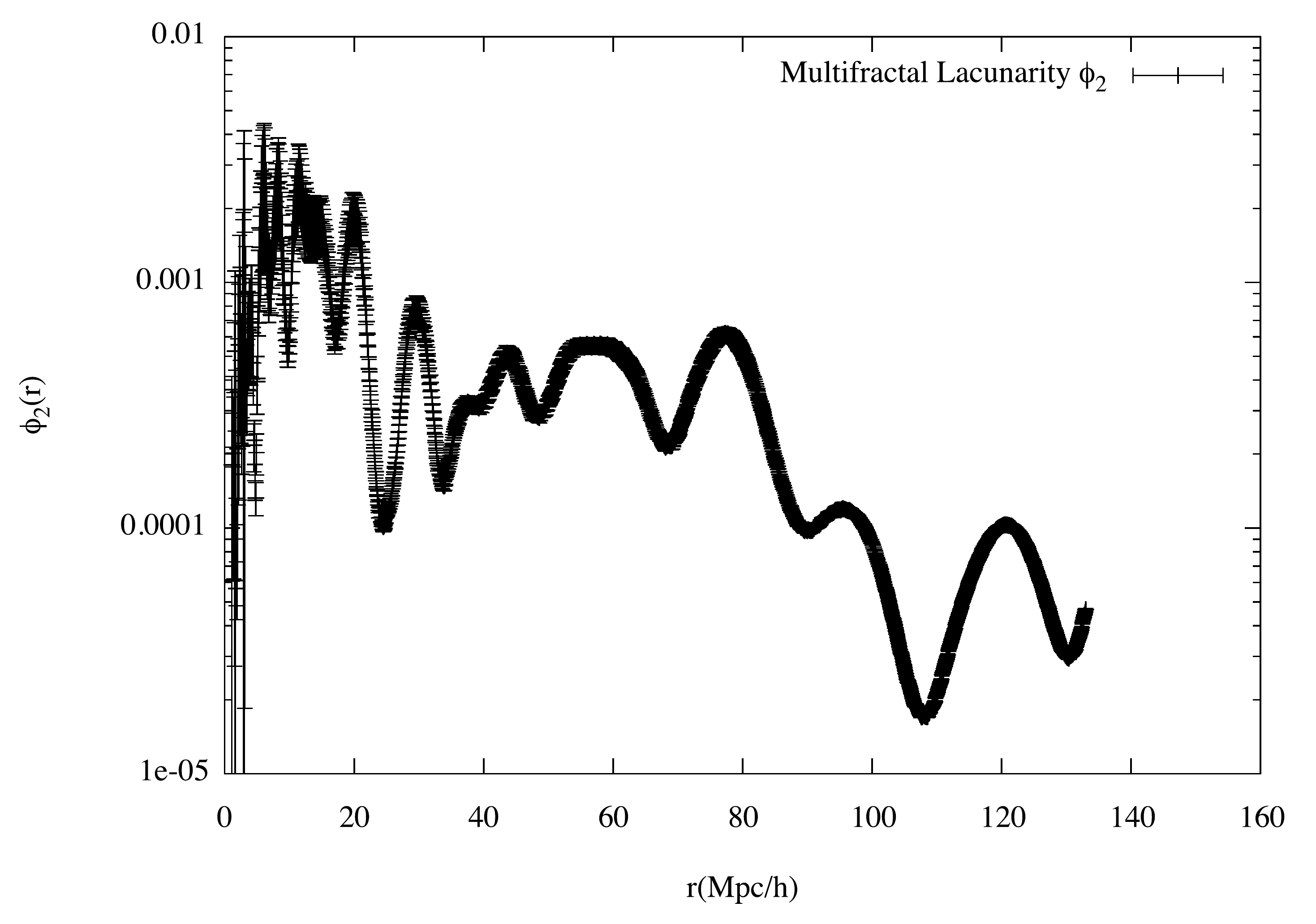}
	\includegraphics[width=0.45\linewidth,clip]{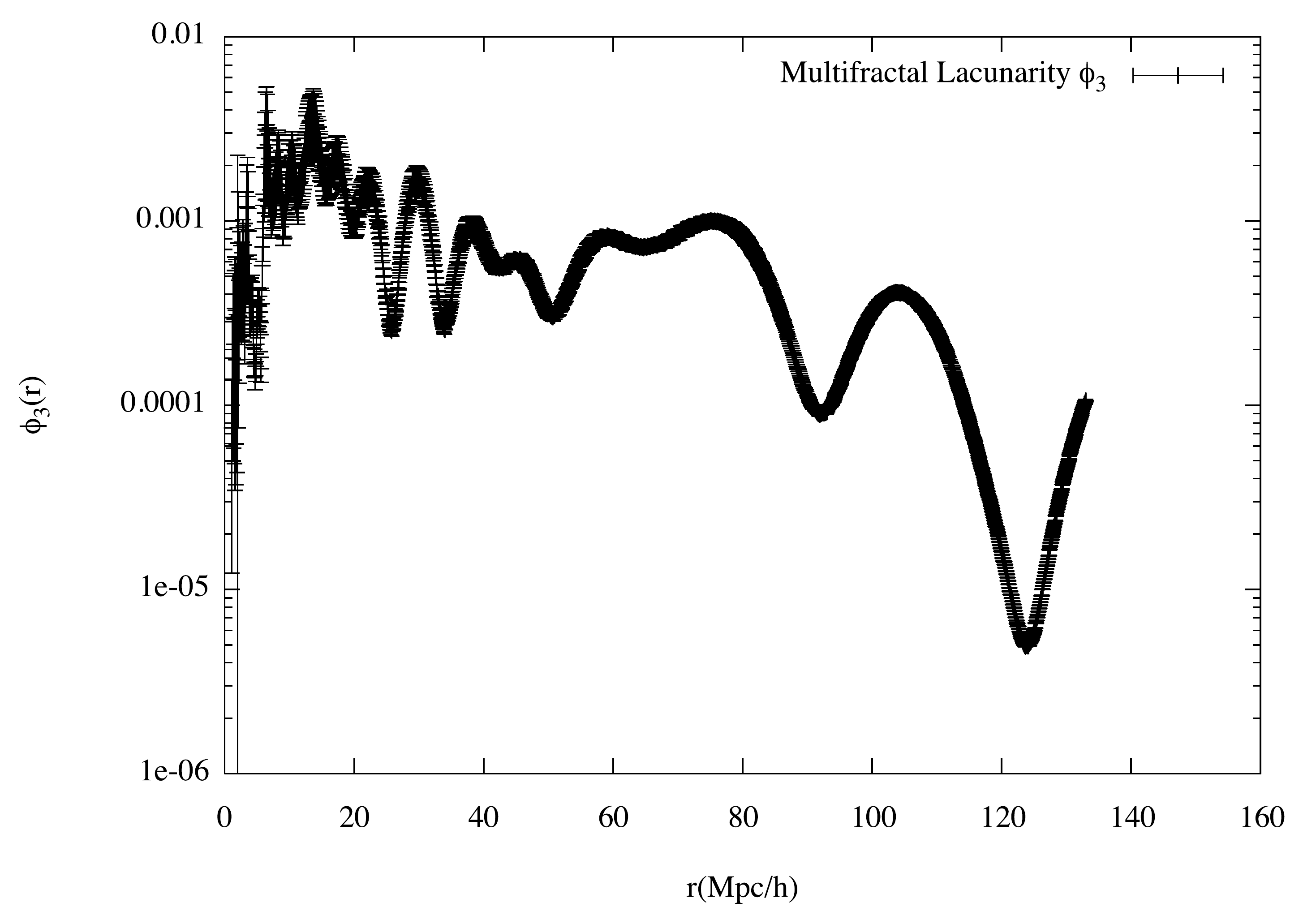}
	\includegraphics[width=0.45\linewidth,clip]{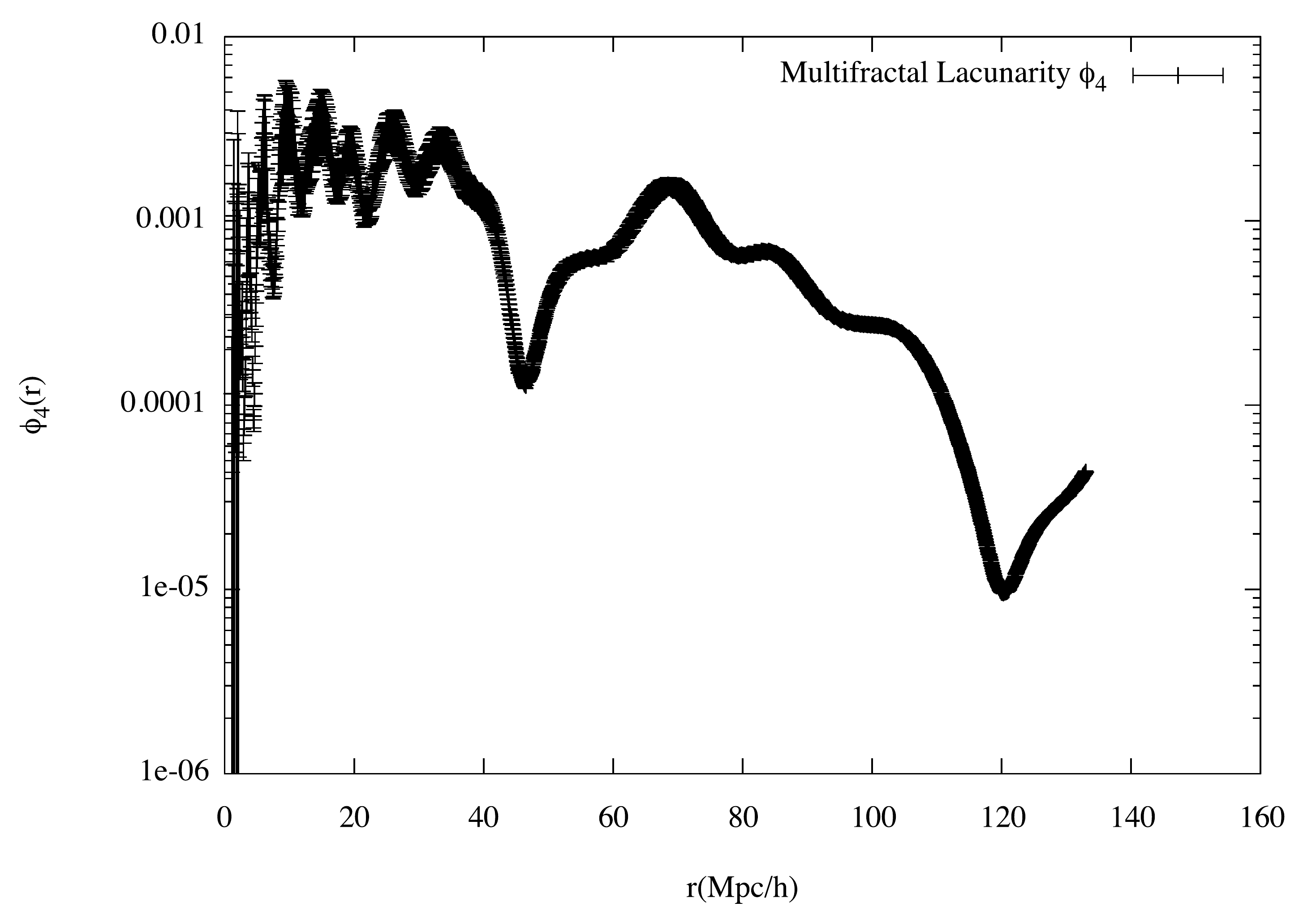}
         \includegraphics[width=0.45\linewidth,clip]{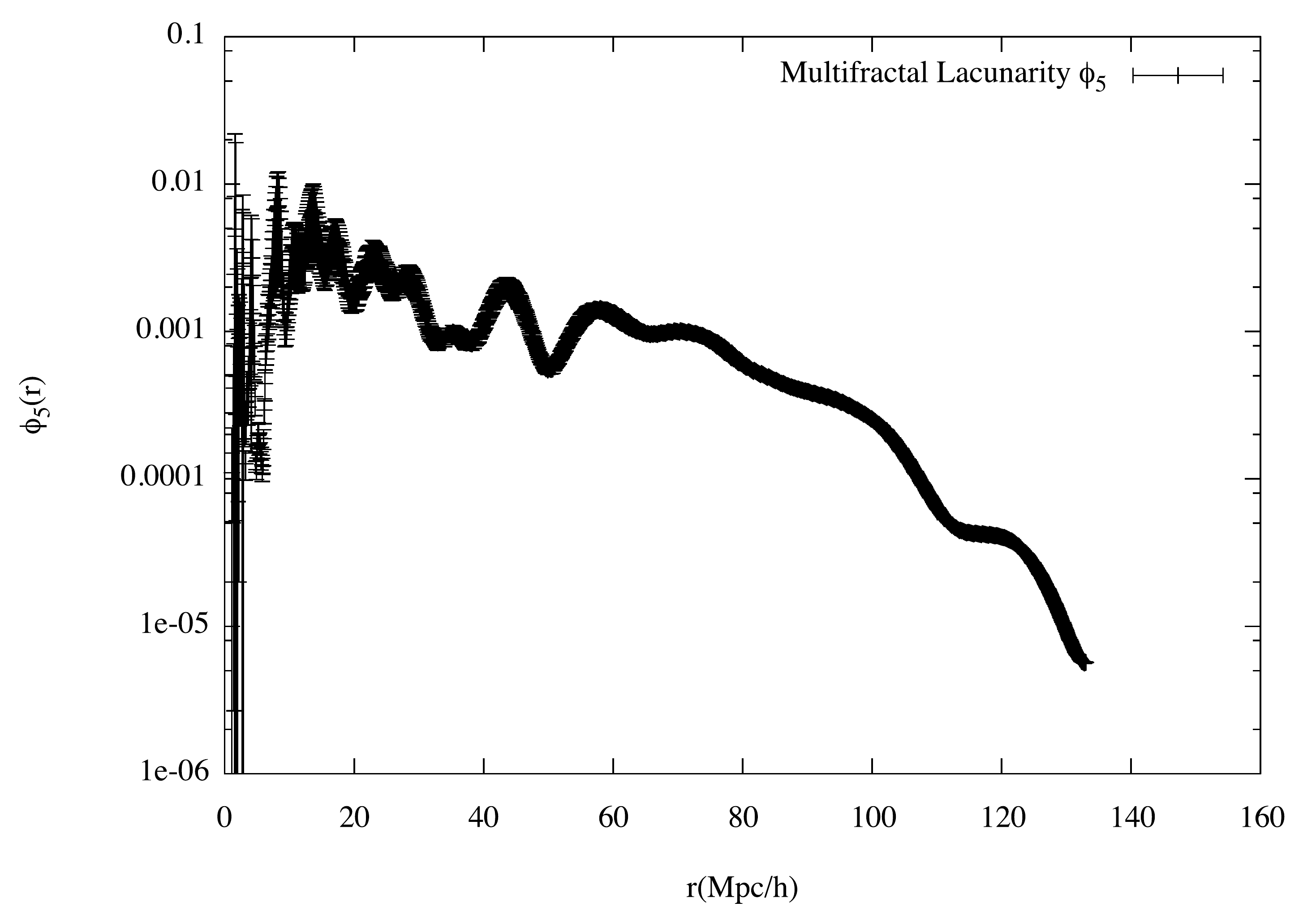}
         \includegraphics[width=0.45\linewidth,clip]{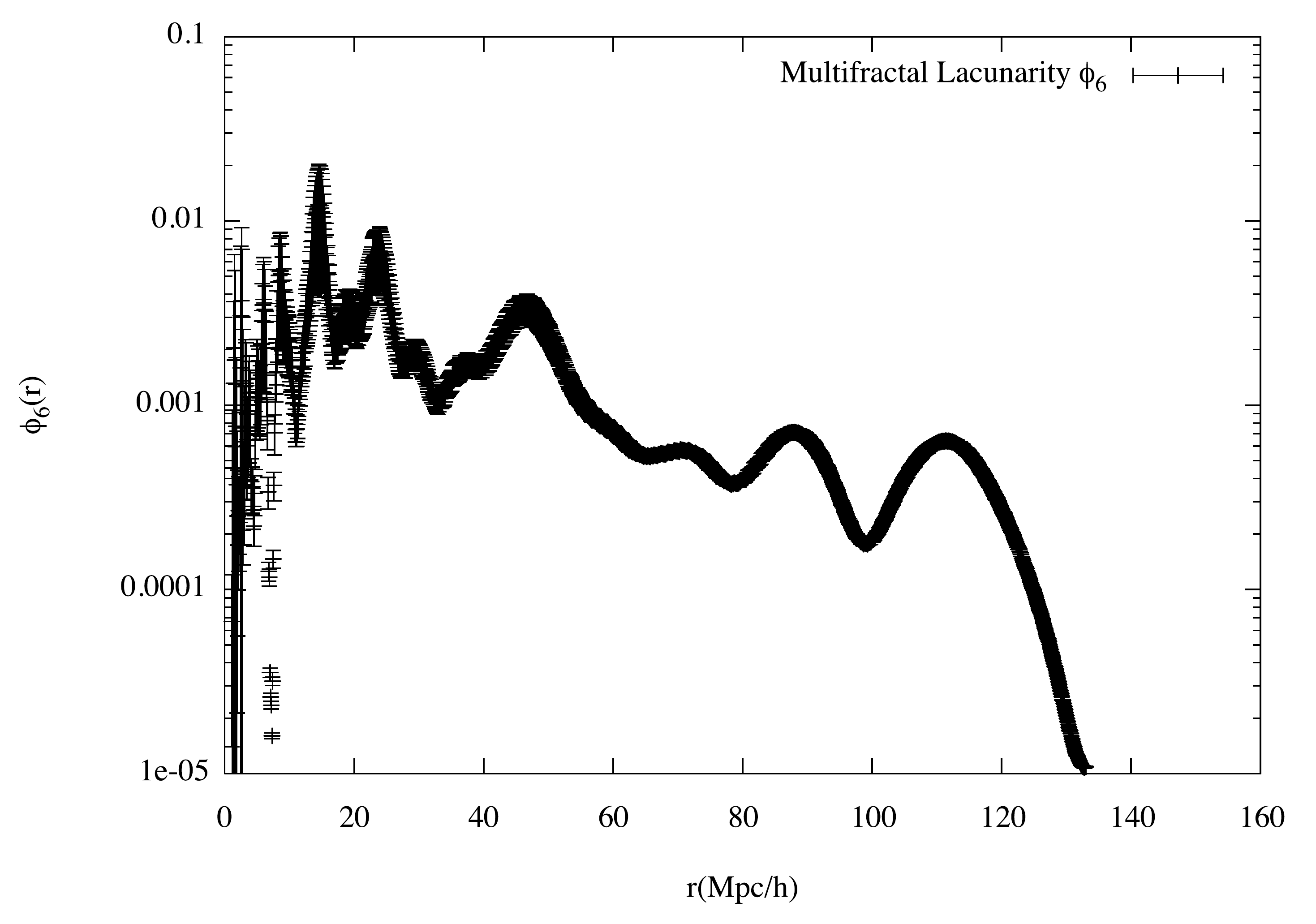} 
         \caption{Multi-lacunarity spectrum $\Phi_q(r)$ in logscale \textit{vs}  radial distance $r$ for high density environments $q \geqslant 1$  from the Millennium dark matter haloes distribution. Oscillatory behaviour modulated by  a decreasing function. $ \pm1\sigma$ error bars are shown, with 0.99 confidence level in t-student test}
	\label{Phiqr}
\end{figure*}

\section{Discussion}

We confirm the dependence of fractal dimension with the radial distance; The most of the previous works reported only a single fractal dimension value to frame the clustering behaviour of large-scale mass \citep{1998A&A...339L..85D, 1999dmap.conf..336S, 2005A&A...443...11J}. The majority of recent works focus on the radial distance dependency of the fractal dimension and the multi-fractal analysis of large scale clustering in the Universe, for example: \citet{2002PhyD..168..404R} describe the scaling properties of large-scale structures in the Universe and the transition to homogeneity around 500 Mpc/h,  \citet{2005BASI...33....1S} found that the universe is homogeneous over scales larger than about 80 to 100Mpc/h,  \citet{2008MNRAS.390..829B} observed that the fractal dimension makes a rapid transition to values close to the physical space dimension at scales between 40 and 100 Mpc, while \citet{2009MNRAS.399L.128S} determinate that the Millennium simulation exhibit a transition to homogeneity at around 70 Mpc/h. In our case, the analysis of The Millennium Simulation confirms fractal behaviour (in the average mass-radius relation) at scales smaller than at least 100 Mpc/h with homogeneity transition around 110 Mpc/h for the dark matter haloes distribution.

On the other hand, we observe limitations in the application of  the average mass-radius fractal dimension and the multi-fractal dimension spectrum to radial distances larger than 180 Mpc/h. We detect spurious homogenisation effects from these calculations as the radial distance from the centres approaches the edges of the simulation. This effect is reported in the literature \citep{1989A&A...219....1W, 1996PhyA..226..195S}. As stated above, due to  this edge effect it is necessary to limit the maximum radius to which the fractal dimension can be calculated within the simulation. In our case the average mass-radius dimension, the lacunarity, the multi-fractal dimension and the lacunarity spectrum have to be limited to maximum radial distances of 160 Mpc/h, where the fractal dimension does not exceed the value of $3 \pm1\sigma$. 

The behaviour of mass-radius fractal dimension can be divided in two regions clearly observed: a fast-growing region from each centre up to $\approx$ 20 Mpc/h in radial distance, followed by a slow-growing region to achieve homogeneity at radial distances around 100 Mpc/h. Above this distance the search for the scale of transition to homogeneity using the lacunarity and the multi-fractal treatment should focus.

The calculation of lacunarity as complement in the determination of mass-radius fractal dimension, in addition to confirm the process of increasing the homogeneity (Decreasing lacunarity values), is capable to detect heterogeneity oscillations with local increase of lacunarity. This behaviour indicates alternation between voids (lacunarity high values) and concentrations of matter (lacunarity low values) and was reported earlier by \citet{1997Chaos...7...82P}, \citet{1997PhLA..228..351S}, \citet{1998NYASA.867..258M}. Furthermore, the high values of lacunarity at radial distances smaller than 60 Mpc/h confirm the existence of voids near each centre, followed by oscillatory lacunarity behaviour (gaps and high concentration of dark matter alternatively). 

The multi-fractal spectrum calculation exhibits less dispersion than that found on the average mass-radius fractal dimension and its corresponding lacunarity, as seen in the behaviour of the $\chi^2$ test. This gives more confidence on the subsequent analyses. The results for structure parameter values $q < 1$, show an excessive increase of the fractal dimension for radial distances smaller than 40 Mpc/h, reaching values that overcome the physical dimension of space (unphysical values), followed by a tendency to homogeneity for voids at larger radial distances. In the work of  \citet{2009MNRAS.399L.128S}, on transition to homogeneity in the Millennium Simulation data,  it is shown that for $q = - 2$ fractal dimension values larger than $3$ appear not only in the region below 40 Mpc/h, but also a radial distances between 70 Mpc/h and 90 Mpc/h. 

For $q \geqslant 1$ we observe a slower growth of the multi-fractal dimension as the parameter structure increases, indicating a heterogeneity growth as hierarchical clustering increases. We find the possible homogeneity transition located in radial distances between 100 Mpc/h and 120 Mpc/h . This result has be corroborated by evaluating the multi-fractal dimension as a function of structure parameter $q$, where a homogeneity transition around 120 Mpc/h is detected.  This homogeneity transition scale is larger than those reported in similar works such as  \citet{2007ApJ...658...11G}, \citet{2009MNRAS.399L.128S}, but lower than the reported by \citet{2010MNRAS.405.2009Y}. Other works on galaxy surveys did not report any homogeneity transition (e.g. \citet{2008EPJB...64..615S} and  \citet{2011ARep...55..324V}). For this reason we propose to use the spectrum of lacunarity as another homogeneity transition indicator.

The lacunarity spectrum presents a similar behaviour to that found in the lacunarity calculations based in the average mass-radius dimension.  An initial increase in the heterogeneity, i.e., great presence of voids in regions close to the centres (below 20 Mpc/h) followed by a decreasing function which module oscillations with lacunarity values alternation. This shows that the clustering of dark matter haloes in the Millennium Simulation have low dense regions followed by regions of high concentration of dark matter haloes. Furthermore, the lacunarity spectrum points to a homogeneity transition (minimum lacunarity values) for radial distances between 100 Mpc/h and 130 Mpc/h,  confirming the region where the cross to homogeneity was found in the multi-fractal spectrum.

\section{Conclusions}

In this paper we have studied the distribution of  the Millennium Simulation dark matter haloes from a multi-fractal viewpoint. Our results provide a complementary method on the detection of the homogeneity transition scale based on the lacunarity concept. Our main conclusions are as follows.

The dark matter haloes clustering in the $\Lambda$CDM Millennium Simulation have a radial distance dependency divided in two regions. First a region with a fast dimensional growth which starts with great variability in regions close to the centres,  around 20 Mpc/h, a symptom of inhomogeneity at short distances. A second region where the clustering of dark matter shows evidence of slower dimensional growth arriving to homogeneity with a progressive decrease in the dispersion. Similar results are reported by \citet{0295-5075-71-2-332, 2007ApJ...658...11G} in his simulation fractal analysis. Therefore, the use of multi-fractal analysis appears indispensable because of the complexity in the clustering of the dark matter haloes. 

In the determination of the fractal mass-radius dimension and multi-fractal spectrum it is necessary to consider the spurious homogenisation effects introduced as the radial distance approaches the edges of the volume to analyse. In our case when it comes near the edges, the number of possible centres are being limited (fewer centres from which to estimate the relationship); Besides, the mass-radius fractal dimension starts to exceed the spatial dimension that the clustering of dark matter haloes is embedding, this can lead to confusing behaviour over the cross to homogeneity. \citep{1989A&A...219....1W, 2002SPIE.4847...86M, citeulike:2502783}. For this reason the maximum radial distance for the fractal analysis should be limited in the calculation, according to the sample spatial size.

For the case of low-density environments, $q < 1$, it is remarkable the coherence between high dimension values for voids below $r \approx 20$ Mpc/h, the first region of fast growing in the multi-fractal dimension spectrum for  $q \geqslant 1$, and the lacunarity spectrum for the same spatial region. This is an evidence of complexity due to self-gravitation at short distances.  

In the calculation of the multi-fractal spectrum, we show the homogeneity cross at depths between 100 Mpc/h and 120 Mpc/h, without exceeding the physical dimension of space. The cross to homogeneity is more precisely located by the relation between the multi-fractal dimension and the structure parameter $q$, which shows the cross to homogeneity at 120 Mpc/h radial distance. The spectrum of lacunarity confirms this result, every lacunarity function points the homogeneity scale in the same spatial region. 

The lacunarity spectrum for every structure parameter in overdenses environments $q \geqslant 1$ reveals regions with relative maxima, allowing us to detect spatial regions where voids are forming inside the clustering of dark matter haloes. The transition of homogeneity for every structure parameter is founded beyond the 100 Mpc/h. Insofar that the structure parameter increases, the minimum value of lacunarity is located farther from the centre, in the same manner that the scale of homogeneity transition is increasing in the multi-fractal spectrum.

\section*{Acknowledgements}
 C. A. Chac\'on acknowledges doctoral financial support from the Universidad Distrital Francisco Jos\'e de Caldas. The authors thank the anonymous referee for the careful reading and the valuable comments and suggestions to improve this paper.
  
\bibliography{my_citations}

\label{lastpage}

\end{document}